\documentclass[%
twocolumn,
superscriptaddress,
 amsmath,amssymb,
 aps,prc,
 floatfix,
10pt
]{revtex4-1}

\usepackage{natbib}
\usepackage[english]{babel}% Niklas and Nathalia option
\usepackage{amsmath}
\usepackage{amssymb}
\usepackage{graphicx}% Include figure files
\usepackage{dcolumn}% Align table columns on decimal point
\usepackage{bm}% bold math
\newcommand{\aZ}{\alpha Z}
\newcommand{\aZs}{(\alpha Z)^2}
\newcommand{\balpha}{\mbox{\boldmath $\alpha$} }

\newcommand{\ket}[1]{| #1 \rangle}
\newcommand{\bra}[1]{\langle #1 |}
\begin{document}

\preprint{APS/123-QED}
\title{The measurement of the quadrupole moment of $^{185}$Re and $^{187}$Re from the hyperfine structure of muonic X rays}

\author{A. Antognini}
\affiliation{Paul Scherrer Institut, Villigen, Switzerland}
\affiliation{Institut f\"ur Teilchen- und Astrophysik, ETH Z\"urich, Switzerland}
\author{N. Berger}
\affiliation{PRISMA+Cluster of Excellence and Institute of Nuclear Physics, Johannes Gutenberg Universit\"at Mainz, Germany}
\author{T.E. Cocolios}
\affiliation{KULeuven, Instituut voor Kern-en Stralingfysica, B-3001 Leuven, Belgium}
\author{R. Dressler}
\affiliation{Paul Scherrer Institut, Villigen, Switzerland}
\author{R. Eichler}
\affiliation{Paul Scherrer Institut, Villigen, Switzerland}
\author{A. Eggenberger}
\affiliation{Institut f\"ur Teilchen- und Astrophysik, ETH Z\"urich, Switzerland}
\author{P.~Indelicato}
\affiliation{LKB Paris, France}
\author{K-P. Jungmann}
\affiliation{VSI, University of Groningen, The Nederlands}
\author{C. H. Keitel}
\affiliation{Max-Planck-Institut f\"ur Kernphysik, Heidelberg, Germany}
\author{K. Kirch}
\affiliation{Paul Scherrer Institut, Villigen, Switzerland}
\affiliation{Institut f\"ur Teilchen- und Astrophysik, ETH Z\"urich, Switzerland}
\author{A. Knecht}
\affiliation{Paul Scherrer Institut, Villigen, Switzerland}
\author{N. Michel}
\affiliation{Max-Planck-Institut f\"ur Kernphysik, Heidelberg, Germany}
\author{J. Nuber}
\affiliation{Paul Scherrer Institut, Villigen, Switzerland}
\affiliation{Institut f\"ur Teilchen- und Astrophysik, ETH Z\"urich, Switzerland}
\author{N.~S.~Oreshkina}
\email{natalia.oreshkina@mpi-hd.mpg.de}
\affiliation{Max-Planck-Institut f\"ur Kernphysik, Heidelberg, Germany}
\author{A. Ouf}
\affiliation{PRISMA+Cluster of Excellence and Institute of Physics, Johannes Gutenberg Universit\"at Mainz, Germany}
\author{A. Papa}
\affiliation{Paul Scherrer Institut, Villigen, Switzerland}
\affiliation{Department of Physics, Universit\`a di Pisa, Italy}
\author{R. Pohl}
\affiliation{PRISMA+Cluster of Excellence and Institute of Physics, Johannes Gutenberg Universit\"at Mainz, Germany}
\author{M. Pospelov}
\affiliation{University of Victoria, Canada}
\affiliation{Perimeter Institute, Waterloo, Canada}
\author{E. Rapisarda}
\email{elisa.rapisarda@psi.ch}
\altaffiliation[Also at ]{e.rapisarda@iaea.org}%Lines break automatically or can be forced with \\
\affiliation{Paul Scherrer Institut, Villigen, Switzerland}
\author{N. Ritjoho}
\affiliation{Paul Scherrer Institut, Villigen, Switzerland}
\affiliation{Institut f\"ur Teilchen- und Astrophysik, ETH Z\"urich, Switzerland}
\author{S. Roccia}
\affiliation{CSNSM, Universit\`e Paris Sud, CNRS/IN2P3, Orsay Campus, France}
\affiliation{Institut Laue-Langevin, CS 20156 F-38042 Grenoble Cedex 9, France}
\author{N. Severijns}
\affiliation{KULeuven, Instituut voor Kern-en Stralingfysica, B-3001 Leuven, Belgium}
\author{A. Skawran}
\affiliation{Paul Scherrer Institut, Villigen, Switzerland}
\affiliation{Institut f\"ur Teilchen- und Astrophysik, ETH Z\"urich, Switzerland}
\author{S. M. Vogiatzi}
\affiliation{Paul Scherrer Institut, Villigen, Switzerland}
\affiliation{Institut f\"ur Teilchen- und Astrophysik, ETH Z\"urich, Switzerland}
\author{F. Wauters}
\affiliation{PRISMA+Cluster of Excellence and Institute of Nuclear Physics, Johannes Gutenberg Universit\"at Mainz, Germany}
\author{L. Willmann}
\affiliation{VSI, University of Groningen, The Nederlands}

%\affiliation{%

%\affiliation{
%\affiliation{
%\affiliation{%

\date{\today}% It is always \today, today,

\begin{abstract}
The hyperfine splitting of the $5g \rightarrow 4f$ transitions in muonic $^{185,187}$Re has been measured using high resolution HPGe
detectors and compared to state-of-the-art atomic theoretical predictions. The spectroscopic quadrupole moment has been extracted using 
modern fitting procedures and compared to the values available in literature obtained from muonic X rays of natural rhenium. 
The extracted values of the nuclear spectroscopic quadrupole moment are 2.07(5) barn and 1.94(5)~barn, respectively for 
$^{185}$Re and $^{187}$Re.\\
This work is part of a larger effort at the Paul Scherrer Institut towards the measurement of the nuclear charge radii
 of radioactive elements. 

%\begin{description}
%\end{description}
\end{abstract}

\pacs{32.30.Rj, 32.10.Fn, 36.10.Ee, 21.10.Ky} % PACS, the Physics and Astronomy
\keywords{Suggested keywords}%Use showkeys class option if keyword
\maketitle

%\tableofcontents

\section{\label{sec:intro} Introduction}
%\blinddocument

It is well known that muonic X rays can be used as a sensitive means to determine the charge radius of a nucleus. 
Moreover if the hyperfine structure (hfs) can be resolved, then the distribution of the magnetic dipole (MD) 
and electric quadrupole (EQ) moments in the nucleus can be investigated as well. All stable elements and few unstable elements
have been studied by muonic X-ray spectroscopy. Rhenium is the last stable element whose nuclear charge radius has not been
measured with muonic X rays~\cite{Angeli2013}. Since Re is a strongly-deformed nucleus, the muonic X-ray spectrum is complicated by 
the so-called dynamic hyperfine splitting~\cite{Jac54,Wil54}. This effect is particularly sizeable in muonic atoms and is due to the fact that
the quadrupole interaction between muon and nucleus has non-vanishing off-diagonal elements which link the ground state and
low-lying excited states of the nucleus. The effect leads to a mixing of the nuclear states due to the similar energy scale between 
the atomic binding energies and the nuclear excitation energies resulting in a dynamic hyperfine splitting even for nuclei
which have zero spin in the ground state where no hfs is to be expected. As a result of the dynamic 
hyperfine splitting, the extraction of the nuclear charge parameters from the $2p\rightarrow 1s$ transition in 
deformed nuclei requires a more elaborated analysis compared to spherical nuclei as shown in~\cite{Ack66,McKee69,tanaka1984} 
and references therein. 

The only existing measurement of muonic X rays of rhenium was performed on a natural rhenium target and aimed at the 
extraction of the spectroscopic quadrupole moment from the analysis of the hyperfine splitting of the 
$5g\rightarrow 4f$ transitions~\cite{Kon81}. \\
The full muonic X-ray spectrum of isotopically pure targets of $^{185}$Re and $^{187}$Re has been recently measured by the 
muX collaboration at the Paul Scherrer Institut (PSI) for the first time, with the aim to extract the main properties
of the nuclear charge distribution from muonic spectroscopy, which are still missing in literature. 
In this paper we present the analysis of the  hyperfine splitting of the $5g \rightarrow 4f$ muonic transitions yielding
the spectroscopic quadrupole moment of $^{185,187}$Re.
The analysis of the $2p\rightarrow 1s$ and $3d\rightarrow 2p$ muonic transitions and the extraction of the nuclear charge radius 
will be reported elsewhere. \\
The muonic X-ray spectra of the two isotopically pure rhenium targets, 
a major improvement over the analysis presented in Ref.~\cite{Kon81},  analysed with state-of-the-art theoretical predictions and 
fitting procedures have shown that the
fit of the hfs of the $5g\rightarrow 4f$ transitions, and consequently the extracted value of the quadrupole moment, 
is very sensitive to the inclusion of  weaker muonic transitions not included in the analysis of Ref.~\cite{Kon81}. 
Particular care has also been taken on the determination of the peak shape of the germanium detectors
from muonic X-ray data, while in~\cite{Kon81} off-line calibration runs with sources were used.
The measurements reported in this article are part of a larger effort going on at PSI (muX project) to perform 
muonic atom spectroscopy on radioactive elements (usually available only in  microgram quantities) aiming, 
as first test cases, at the precise measurement of the nuclear charge radius of $^{226}$Ra and $^{248}$Cm~\cite{Rap18,Ska18}. 

Section~\ref{sec:theory} reviews the theory of the muonic atoms in order to establish the notation used in the hfs formalism and
also includes a discussion of the various corrections to the energy levels beyond the predictions 
of the Dirac equation. Section~\ref{sec:setup} describes the apparatus used in 
obtaining the X-ray spectra, Section~\ref{sec:calibration} describes the data-reduction methods and also includes the $^{208}$Pb 
muonic X-ray results which where used as a calibration standard for the $^{185,187}$Re spectra. Section~\ref{sec:results} details 
the fit of the $5g \rightarrow 4f$ transitions of the $^{185,187}$Re spectra using the hfs formalism of Section~\ref{sec:hfs} and the extraction 
of the nuclear quadrupole moments.

%\subsection{\label{sec:level2}Second-level heading: Formatting}
%\subsubsection{Wide text (A level-3 head)}
%\paragraph{Note (Fourth-level head is run in)}

\section{\label{sec:theory}Theory}
\subsection{\label{sec:finestructure}Fine structure}
In order to predict the transition energies and probabilities as well as their dependence on the nuclear quadrupole moment theoretically, 
the bound muon is described as a Dirac particle. As the mass of the muon $m_\mu$ is about 207 times larger than the electron's mass $m_e$, 
the energy scale for muonic atoms is a factor $\sim$207 larger than regular electronic atoms.
The Bohr radius of the muon is smaller than the one of the electron by the same factor, which leads to a significant enhancement of nuclear effects.

The bound muon is described by the Dirac equation
\begin{equation}
\label{eq:dirac_muonic}
[\balpha \cdot \mathbf{p} + \beta m_\mu + V_{\text{nucl}}(r) ] \left|n\kappa m\right> = E_{n\kappa}\left|n\kappa m\right>,
\end{equation}
where $\balpha, \beta$ are the Dirac matrices, $V_{\text{nucl}}$ is the electrostatic potential caused by the nuclear charge distribution, 
and $E_{n\kappa}$ and $\left|n\kappa m\right>$ are muonic energies 
and wavefunctions, correspondingly. Here $n$ stands for the principal quantum number, while
the relativistic angular quantum number $\kappa$ is introduced as a bijective function of the orbital angular 
momentum $l$ and the total muon angular momentum $j$ as $\kappa = (-1)^{j+l+1/2}(j+1/2)$, and $m$ is the $z$ component of $j$.
For a spherically symmetric potential, the radial components $G_{n\kappa}(r), F_{n\kappa}(r)$ and the
angular part $\Omega_{\pm \kappa m}$ can be separated, and therefore the solution can be written as~\cite{greiner2000}
\begin{equation}
\left|n\kappa m\right> = \frac{1}{r}\binom{G_{n\kappa}(r)\Omega_{\kappa m}(\mathbf{n})}
	{iF_{n\kappa}(r)\Omega_{-\kappa m}(\mathbf{n})},
\end{equation}
The angular part of the wave function is described by spherical spinors $\Omega_{\kappa m}$, and the radial wave functions are normalised with an integral
\begin{equation}
\int_0^\infty dr[G_{n\kappa}^2(r) + F_{n\kappa}^2(r)] = 1.
\end{equation}

For a Coulomb potential $V_{\text{nucl}}^{\rm C} (r) = -\aZ/r$, Eq.~\eqref{eq:dirac_muonic} can be solved analytically and gives 
the well-known formula for the Dirac-Coulomb energies
\begin{equation}
E_{n\kappa}^{\rm C} = \left[ 1 + \frac{\aZs}{\left(n-|\kappa|+\sqrt{\kappa^2 - \aZs}\right)^2} \right]^{-1/2}
\end{equation}
where $\alpha$ is the fine-structure constant and $Z$ the nuclear charge.
However, predictions of the muonic spectra have to include the finite size of the nucleus already in the Dirac equation. 
The deformed Fermi distribution
\begin{equation}
\rho_{ca\beta}(\mathbf{r})=\cfrac{N}{1+\exp [(r-c[1+\beta \text{Y}_{20}(\vartheta)])/{a}]}
\label{eq:deffermi}
\end{equation}
has proven to be very successful in the description of the level structure of heavy muonic atoms, see e.g.~\cite{hitlin1970,tanaka1984,tanaka1984_2}, 
and is also used in this work. Here, $a$ is the skin thickness parameter, $c$ the half-density radius, $\beta$ the deformation parameter, 
$N$ a normalisation constant and $Y_{20}$ the spherical harmonics. The corresponding spherically symmetric part of the nuclear potential is
\begin{equation}
V_{\rm{nucl}}(r)= -\alpha \int\rm{d}^3\mathbf{r}^{\prime}\frac{\rho_{ca\beta}(\mathbf{r}^\prime)}{\max(r,r^\prime)}.
\label{eq:sphepot}
\end{equation}
It has been shown, that $a=t/(4\,\text{log}3)$, with $t=2.30\,\text{fm}$, is a good approximation for most of the nuclei~\cite{Beier2000}. 
Then, $c$ and $\beta$ are chosen such that the root-mean-square radius $r_{\rm RMS}$ of the distribution agrees with the literature value~\cite{Angeli2013} 
and the quadrupole moment agrees with a given value, which is obtained by fitting to the experimental data as described in Section~\ref{sec:fit}. 
The connection between the charge distribution of Eq.~\eqref{eq:deffermi} and the spectroscopic quadrupole moment is
\begin{equation}
Q = \frac{2I(2I-1)}{(I+1)(2I+3)}\int\rm{d}^3\mathbf{r}^\prime \,r^{\prime 2}\rho_{ca\beta}(\mathbf{r}^\prime)
\text{P}_2(\cos\vartheta^\prime),
\end{equation}
where $I$ is the nuclear angular momentum number and $\text{P}_l(x)$ are the Legendre polynomials.

With the potential of Eq.~\eqref{eq:sphepot}, Eq.~\eqref{eq:dirac_muonic} can be solved only numerically. For this purpose the dual-kinetic-balance 
method~\cite{Shabaev2004} has been used in this work. For the muon in the $1s$ state the binding energy including finite-size effect is almost 50\%
smaller than the value $E_{n\kappa}^{\rm C}$ assuming a point Coulomb potential.  For the $4d$ states the reduction is on a level of 0.1\%, and even 
smaller for the $4f$ states. 

The order-$\alpha$ quantum electrodynamics contributions are the self-energy (SE) and the vacuum polarisation (VP) corrections. 
For atomic electrons they are usually of the same order of magnitude.
For muons, however, the VP correction is much larger as the virtual electron-positron pair production is less suppressed due to 
their low mass compared to the muon's mass \cite{BorieRinker1982}. 
The dominant VP contribution (first order in $\alpha$ and $\aZ$) is called Uehling correction, and can be described by the potential~\cite{Elizarov2005}
\begin{align}
&V_{\text{Uehl}}(r)=-\alpha \frac{2\alpha}{3\pi}{\int\text{d}^3\mathbf{r}^{\prime}\,\rho_{ca\beta}(\mathbf{r}^\prime)}\int_1^\infty \text{d}t\,\left( 1+\frac{1}{2t^2} \right)\nonumber\\
&\times\frac{\sqrt{t^2-1}}{t^2} \frac{\text{exp}(-2m_e|r-r^\prime|t)-\text{exp}(-2m_e(r+r^\prime)t)}{4m_er t}.
\label{eq:uehl}
\end{align}
This potential can be directly included into the Dirac equation of Eq.~\eqref{eq:dirac_muonic} by adding it to $V_{\rm{nucl}}(r)$, 
therefore directly accounting all iterations \cite{indelicato2013} of the Uehling potential into the muonic binding energies.
In the same way, the higher-order contributions to the VP correction, namely the Wichmann-Kroll (order $\alpha(\alpha Z)^3$) 
potential~\cite{wichmann1956,Fullerton1976} in the 
point-like approximation and the K\"allen-Sabry (order $\alpha^2(\alpha Z)$) 
potential~\cite{kallen1955} for a spherically symmetric nuclear charge distribution were included in the Dirac equation, using the 
expressions from~\cite{indelicato2013}.
Since both the Wichmann-Kroll and the K\"allen-Sabry  corrections to the energy levels of muonic atoms are small, 
the neglected nuclear model dependence was estimated to be insignificant. 

The recoil correction, i.e., the effect of finite nuclear mass and the resulting motion of the nucleus, was accounted following the approach used 
in Refs.~\cite{friar1973,BorieRinker1982,michel2017}. 

The effect of the surrounding electrons on the binding energies of the muon, commonly referred to as electron screening, was estimated 
following Ref.~\cite{vogel1973,michel2017} by calculating an effective screening potential from the charge distribution of the electrons and 
using this potential in the Dirac equation for the muon. 
The atomic electrons primarily behave like a charged shell around the muon and the nucleus; thus every muon level is mainly shifted by a constant 
term, which is not observable in the muonic transitions. 
The main contribution to the screening potential comes from the $1s$ electrons, since their wave functions have the largest overlap with the 
muonic wavefunctions.

The results of our calculations for rhenium for the total binding energies and for the individual contributions are presented in Table~\ref{tb:fine_str}. 

\begin{table}
\caption{\label{tb:fine_str} Contribution to the binding energy of the muonic rhenium assuming the charge distribution of Eq.~\eqref{eq:deffermi}
with the parameters ${c}{=}{6.3500}\,$fm, ${a}{=}{0.5234}\,$fm, 
${\beta}{=}{0.2343}$, which corresponds to ${Q}{=}{2.21}\,\rm{barn}$ and 
$r_{\rm{RMS}}{=}{\,5.3596}\,\rm{fm}$. For a given nuclear charge distribution, the numerical uncertainties are estimated to be below $1\,\rm{eV}$. 
$E_{n\kappa}^{\rm C}$ are the point-like Dirac-Coulomb binding energies and $\delta E_{\rm{fs}}$ the finite nuclear size correction. 
$\delta E_{\rm{uehl}}$, $\delta E_{\rm{ks}}$, and $\delta E_{\rm{wk}}$ are the corrections due to the Uehling\mbox{-,} 
K\"allen-Sabry-, and Wichmann-Kroll potential, respectively. 
$\delta E_{\rm{screen}}$ is the non-constant part of the screening correction due to the surrounding $1s$ electrons. All energies are in keV.}
\begin{ruledtabular}
\begin{tabular}{c|cccccc}
& $E_{n\kappa}^{\rm C}$&$\delta E_{\rm{fs}}$ & $\delta E_{\rm{uehl}}$ & $\delta E_{\rm{ks}}$ & $\delta E_{\rm{wk}}$ & $\delta E_{\rm{screen}}$ \\\hline
$4d_{3/2}$ & 1013.125 & -1.175 & 3.547 & -0.067 & 0.026 & -0.062 \\
$4d_{5/2}$ & 1000.021 & -0.478 & 3.374 & -0.065 & 0.024 & -0.064 \\
$4f_{5/2}$ & 1000.021 & -0.004 & 2.930 & -0.064 & 0.021 & -0.048 \\
$4f_{7/2}$ & \phantom{1}993.697  & -0.001 & 2.859 & -0.063 & 0.020  & -0.049 \\
$5f_{5/2}$ & \phantom{1}640.055  & -0.003 & 1.459 & -0.035 & 0.010  & -0.123 \\
$5f_{7/2}$ & \phantom{1}636.806  & -0.001 & 1.425 & -0.034 & 0.010  & -0.125 \\
$5g_{7/2}$ & \phantom{1}636.806  & -0.000 & 1.215 & -0.033 & 0.009  & -0.098 \\
$5g_{9/2}$ & \phantom{1}634.883  & -0.000 & 1.199 & -0.033 & 0.009  & -0.099
\end{tabular}
\end{ruledtabular}
\end{table}

%-------------------------------------------------------------------------------------------

\subsection{\label{sec:hfs}Hyperfine structure}
The hyperfine splitting appears as a result of the interaction of the bound muon with the
magnetic dipole (MD) and electric quadrupole (EQ) moments of the nucleus.
In contrast to the electronic atom, where the MD splitting dominates over the EQ splitting (see e.g. \cite{Korzinin2005}), 
the muonic MD splitting is suppressed because the magnetic moment of the muon is $m_\mu/m_e$ times smaller than the electronic one. 

As the hyperfine splitting mixes the nuclear and muonic quantum numbers, they are not conserved anymore and cannot be used 
for a proper description of the energy levels. 
Therefore, a combined mixed state with total angular momentum $F$ and its projection $M_F$ is introduced as
\begin{equation}
\label{eq:totalState}
\left| FM_F\,I\,n\kappa\right> = \sum_{M_I, m_j}\text{C}^{FM_F}_{IM_I\,jm_j}|IM_I\rangle
|n\kappa m_j\rangle,
\end{equation}
where $\text{C}^{jm}_{j_1m_1\,j_2m_2}$ are the Clebsch-Gordan coefficients.

The diagonal matrix elements of the EQ hyperfine operator $\widehat{H}_{\rm EQ}$~\cite{Korzinin2005,michel2017,michel2018} 
are determined by the formula:
\begin{align}
\label{eq:hquad}
& E_{\rm EQ} = \bra{FM_FIn\kappa}\widehat{H}_{\rm EQ}\ket{FM_FIn\kappa} \\
& = \alpha Q(-1)^{j+I+F} 
\left\{\begin{array}{ccc} j & I & F \\ I & j & 2\end{array}\right\} \nonumber \\
& \times \sqrt{\frac{(2I+3)(2I+1)(I+1)}{4I(2I-1)}} \sqrt{\frac{(2j+3)(2j+1)(2j-1)}{16j(j+1)}} \nonumber \\
& \times \int_0^\infty \left[ G^2_{n\kappa}(r)+F^2_{n\kappa}(r)\right]\frac{F_{\text{QD}}(r)}{r^{3}} \mathrm{d}r. \nonumber
\end{align}
Here, $F_{\text{QD}}$ is the quadrupole distribution function, which describes the deviations from a point-like quadrupole and depends on 
a deformed charge distribution as
\begin{equation}
\frac{Q F_{\rm{QD}}(r)}{r^3}
=
\frac{2I(2I-1)}{(I+1)(2I+3)}
\int\rm{d}^3\mathbf{r^\prime}\,\rho(\mathbf{r}^\prime)\frac{r_<^2}{r_>^3}\rm{P}_2(\cos\vartheta^\prime)
\end{equation} 
where $r_< = \min(r,r^\prime)$ and $r_> = \max(r,r^\prime)$. Similarly, the MD hyperfine splitting can be calculated 
by the formula~\cite{Korzinin2005,michel2017,michel2018}
\begin{align}
\label{eq:hmag}
E_{\rm MD} &= \bra{FM_FIn\kappa}\widehat{H}_{\rm MD}\ket{FM_FIn\kappa} \\
&=\left[ F(F+1)-I(I+1)-j(j+1)\right] \\
&\times\frac{\alpha}{2 m_p}\frac{\mu}{\mu_N}\frac{\kappa}{Ij(j+1)} \nonumber \\
&\times \int_0^\infty G_{n\kappa}(r)F_{n\kappa}(r)\frac{F_{\text{MD}}(r)}{r^2}\mathrm{d}r,\nonumber
\end{align}
with the proton mass $m_p$, nuclear magneton $\mu_N$, nuclear magnetic dipole moment $\mu$, and its distribution function $F_{\rm{MD}}(r)$.
For the simple model of a homogeneous  distribution of the dipole moment inside the nucleus, $F_{\rm{MD}}$ reads
\begin{equation}
\label{eq:bwsimple}
F_{\rm{MD}}(r)=\begin{cases}
\left( \cfrac{r}{R_N} \right)^3 & r \leq R_N\\
1 &r > R_N
\end{cases}.
\end{equation}
where for R$_N$ the nuclear charge radius is commonly used.
In practice, both the electric  $F_{\rm{EQ}}$ and magnetic  $F_{\rm{MD}}$ distribution functions were calculated 
for several nuclear models to estimate the model uncertainty using the values of the nuclear 
magnetic moment $\mu / \mu_N = 3.1871$ for $^{185}$Re and $\mu / \mu_N = 3.2197$ for $^{187}$Re 
~\cite{Stone2016}.

%-------------------------------------------------------------------------------------------

\subsection{Dynamical splitting}
\label{sec:dyn}

For the $2p$ states in heavy muonic atoms, the EQ hyperfine splitting, the fine-structure splitting, and the low-lying nuclear 
rotational band can be on the same energy scale of few hundreds of keV.
This leads to a strong mixing of the muonic and nuclear levels caused by the EQ hyperfine interaction, commonly called dynamic 
hyperfine  splitting~\cite{hitlin1970}.

For the analysis of the transitions from ${n}{=}{5}$ to ${n}{=}{4}$ in this work, the hyperfine splitting is much smaller than the nuclear
transitions between low-lying nuclear states, hence the excited nuclear states do not need to be considered. 
However, there is still a residual mixing of the muonic states of Eq.~\eqref{eq:totalState} due to higher-order hyperfine interaction. 
This can be included by rediagonalisation of the EQ and MD interaction in the considered initial and final states.

For the set of all considered initial/final states, the non-diagonal  EQ and MD matrix elements of the $\widehat{H}_{\rm EQ}$ and $\widehat{H}_{\rm MD}$ operators 
have been calculated~\cite{michel2018,michelphd}. Then, the rediagonalisation has been performed separately for each value of $F$, 
since the MD and EQ interaction are diagonal in $F$. 
After the rediagonalisation, the unperturbed states $\left|FM_F\,I\,n\kappa\right>$ are mixed and can be described as
\begin{equation}
\label{eq:rediagonState}
\left|FM_F,\,i\right> = \sum_{k=1}^d c^{(i)}_k \left| FM_F\,I\,n_k\kappa_k\right>,
\end{equation}
where  $d$ is the number of initial/final states, and the coefficients $c^{(i)}_k$ diagonalise the hyperfine interaction.
The quantum numbers $F$ and $M_F$, describing the total angular momentum of the nucleus-muon system, are still well-defined. 
In this work, the EQ matrix elements were also corrected with the order $\alpha(Z\alpha)$ VP contribution using the 
approach of~\cite{michel2018}.

%-------------------------------------------------------------------------------------------
\subsection{Transition probabilities and line intensities}

The muonic transition rates due to spontaneous emission of a photon between states with defined total angular momentum F
from an initial state $\left|F_iM_i,i_i\right>$ to a final state $\left|F_fM_f,i_f\right>$, summed over the projections $M_i$ and $M_f$ 
(to simplify the formalism $M \equiv M_F$ from now on), are~\cite{johnson2007}
\begin{align}
\label{eq:transitionGeneral}
A^{(\lambda)}_{J}&=\frac{2\alpha (2J+1)(J+1)}{J}\Delta E^{if} \\ 
&\times \sum_{M,M_i,M_f} \left|\left<F_fM_f,i_f\middle|\hat{t}^{(\lambda)}_{JM}\middle|F_iM_i,i_i\right>\right|^2. \notag
\end{align}
Here, $\Delta E^{if}$ is the energy difference between the initial and final state, $J$ is the total angular momentum of the photon 
and $\hat{t}^{(\lambda)}_{JM}$~\cite{johnson2007} is the multipole transition operator. $\lambda=1$ corresponds to an electric transition, whereas $\lambda=0$ stands for a magnetic transition.

In the experimental spectra, the number of counts measured in the peak is proportional to the transition intensities, which are the product of the transition probability 
and the population of the initial states. The transition probability per unit time can be calculated 
{\it ab initio} with Eq.~\eqref{eq:transitionGeneral}. 
In this work, the relative population of the muonic fine structure states within a $l$ state was assumed statistical, i.e. proportional to $j(j+1)$, 
whereas the relative population of the $5g$ and $5f$ states was left as free parameter and determined by fitting the experimental spectra 
(see Section~\ref{sec:hfsanalysis}).

%-------------------------------------------------------------------------------------------

\subsection{Dependence of observables on quadrupole moment}
\label{sec:fit}

After a muon is captured in a highly excited state and starts cascading towards its ground state, there is an intermediate region, 
($n\approx 5$) where finite nuclear size effects are still rather small while the muon is not significantly influenced 
by the surrounding atomic electrons. This intermediate region (in our case ${n}{=}{5}\rightarrow{n}{=}{4}$) is well suited
for the extraction of quadrupole moments~\cite{Dey1979,konijn1979}.

Four fine-structure states $5g_{9/2}, \, 5g_{7/2},\,5f_{7/2},\, 5f_{5/2}$ together with the nuclear ground state with $I=5/2$ define the initial states. 
The energies were calculated as described in Section~\ref{sec:finestructure}, \ref{sec:hfs}, and \ref{sec:dyn}; including finite size effects, 
VP (Uehling, K\"allen-Sabry, Wichmann-Kroll in point-like approximation, quadrupole electronic-loop Uehling), SE, electron screening, and 
recoil effect; with the rediagonalisation of the EQ and MD hyperfine interaction.
The same procedure was repeated for the final states with ${n}{=}{4}$, i.e. $4f_{7/2},\,4f_{5/2},\,4d_{5/2},\,4d_{3/2}$ and $I=5/2$.
The transition probabilities were calculated from each initial to each final state with Eq.~\eqref{eq:transitionGeneral} 
for E1 (${\lambda}{=}{1}$, ${J}{=}{1}$) and M1 (${\lambda}{=}{0}$, ${J}{=}{1}$) transitions, assuming a statistical initial 
population in $M_i$ and $M_f$. 
With this approach, the entire spectrum of interest can be calculated for a given spectroscopic quadrupole moment $Q$.

For the comparison of the theoretical predictions with the measured experimental spectra, the full calculations for each transition 
were performed for several values of the quadrupole moment $Q$ in the proximity of the expected value and a quadratic function 
is fitted for every transition energy and intensity as
\begin{align*}
\Delta E^{if}(Q)&= \Delta E^{if}_0 +  \Delta E^{if}_1 Q+  \Delta E^{if}_2 Q^2,\\
I^{if}(Q)&= I^{if}_0 +  I^{if}_1 Q+  I^{if}_2 Q^2.
\end{align*} 
In this way, the fitting coefficients, in addition to the first-order EQ splitting, contain also the information about MD splitting and higher-order EQ interaction, whereas in Ref.~\cite{Kon81} only the term linear in the quadrupole moment $Q$ was considered. 
The resulting dependencies for the transition energies and for the relative intensities are given in
Table~\ref{tab:table3}, in Table~\ref{tab:table3a}, respectively for $^{185}$Re and $^{187}$Re, and in Table~\ref{tab:table4}.
%%%%%%%%%%%%%%%%%%%%%%%%%%%%%%%%%%%%%%%%%%%%%%%%%%%%%%%%%%%%%%%%%%%%%%%%%%%%%%%%%%%%%%%%%%%%%%%%%%%

\section{\label{sec:expe}Experimental setup and analysis}
\subsection{\label{sec:setup} Setup}
The experiment was performed at the HIPA facility of the Paul Scherrer Institut
and is part of the ongoing muonic X-ray study of radioactive elements.
The negative muon beam was obtained from the decay of pions produced in the collisions of
590 MeV protons on a thick graphite target. The momentum-analysed muon beam
was transported to the $\pi$E1 area and consisted mostly of muons and electrons. The electron
contamination, which can be a source of background, was efficiently
removed using a Wien filter separator placed at around 15 m before the target.
As a result, a high purity negative muon beam could be obtained.
The energy of the muon beam was tuned to a momentum of around 29 MeV/c in order to maximise 
the stopping in the targets.
The typical intensity at the detection setup at the given momentum was in the order of 10$^4$ $\mu^-$ per second.\\
The beam exits the beam line through a 75 $\mu$m thick mylar window and travels in air for around 10~cm 
before being stopped in the target.  
The incoming negative muons and electrons were identified before impinging on the target
by the muon counting detector, a 500~$\mu$m thick plastic scintillator
 with a 6$\times$6~cm$^2$ active area read out by photomultipliers and placed in air in close 
vicinity to the end of the beam line. Given the small thickness, the signals induced by the muons
could be easily separated with a threshold cut from the much smaller signals induced by the electrons.
The muon counting detector was used as start detector for the coincidence measurements (see Section~\ref{sec:results}). 
In addition, at the same position, a second scintillator,  2~mm thick with a  9$\times$9~cm$^2$ active area and 
a central  hole of 45~mm, so that the muon beam was passing through this hole before being stopped in the target, was used
as veto detector to produce anti-coincidence conditions on the muonic X-ray spectra.\\

\begin{figure}[t]
\centering{\includegraphics[scale=0.165]{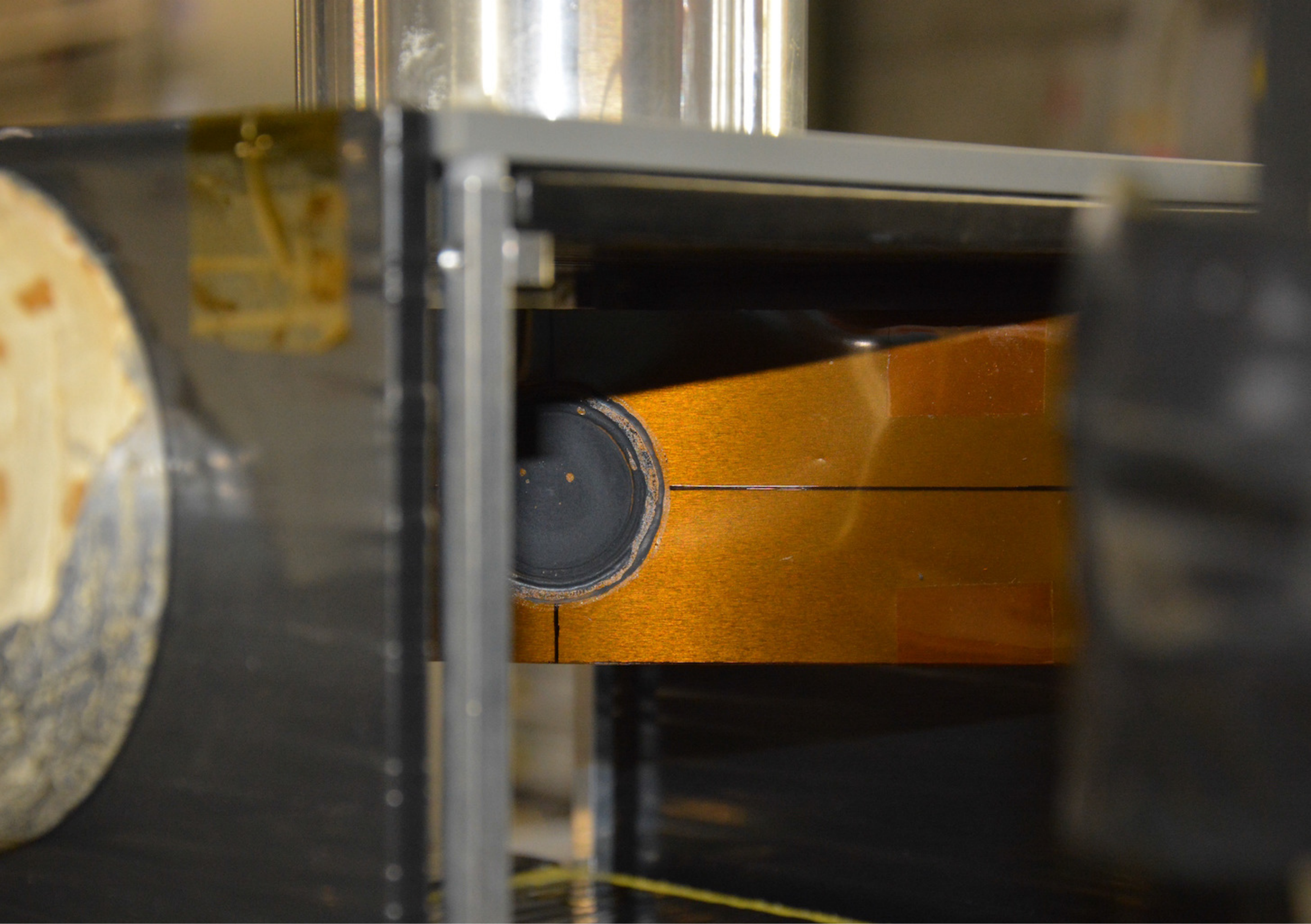}}
\caption{Rhenium target (black disk) glued on a Kapton foil (orange) and mounted on the target holder 
at the centre of the detector arrangement. 
\label{fig01}} 
\end{figure}

Measurements were done with three isotopically pure targets of $^{185}$Re (97.6\%), $^{187}$Re (99.4\%) and 
$^{208}$Pb~(99.6\%). The $^{208}$Pb target was used for the energy calibration and served as a means of 
checking drifts and possible malfunctions. \\The isotopes were purchased in the form of a powder (500~mg) 
in the case of rhenium and in the form of an irregularly shaped ingot (1g) in the case of lead. 
The rhenium powder was first finely ground in a mortar and then mixed with 60 to 70~mg of epoxy on a Kapton foil. 
The mixture was subsequently covered with a Teflon foil and, loaded with some weights, slowly brought into
a disk-like shape of around 30 mm diameter. The lead piece was cold-pressed and hammered into a disk of 40 mm diameter. \\
The targets were then glued onto a Kapton foil and mounted on a PVC frame which was inserted in a target holder 
at 45$^\circ$ with respect to the direction of the beam.
A picture of one of the rhenium  targets mounted on the target holder can be seen in Fig.~\ref{fig01}. 
Typical muon stopping rates were 2500/s for the $^{208}$Pb target and 900/s for the rhenium targets. 
%%%%%%%%%%%%%%%%%%%%%%%%%%%%%%%%%%%%%%%%%%%%%%%%%%%%%%%%%%%%%%%%%%%%%
%%%%%%%%%%%%%%%%%%%%%%%%%%%%%%%%%%%%%%%%%%%%%%

\begin{figure}[t]
\centering{\includegraphics[scale=0.16]{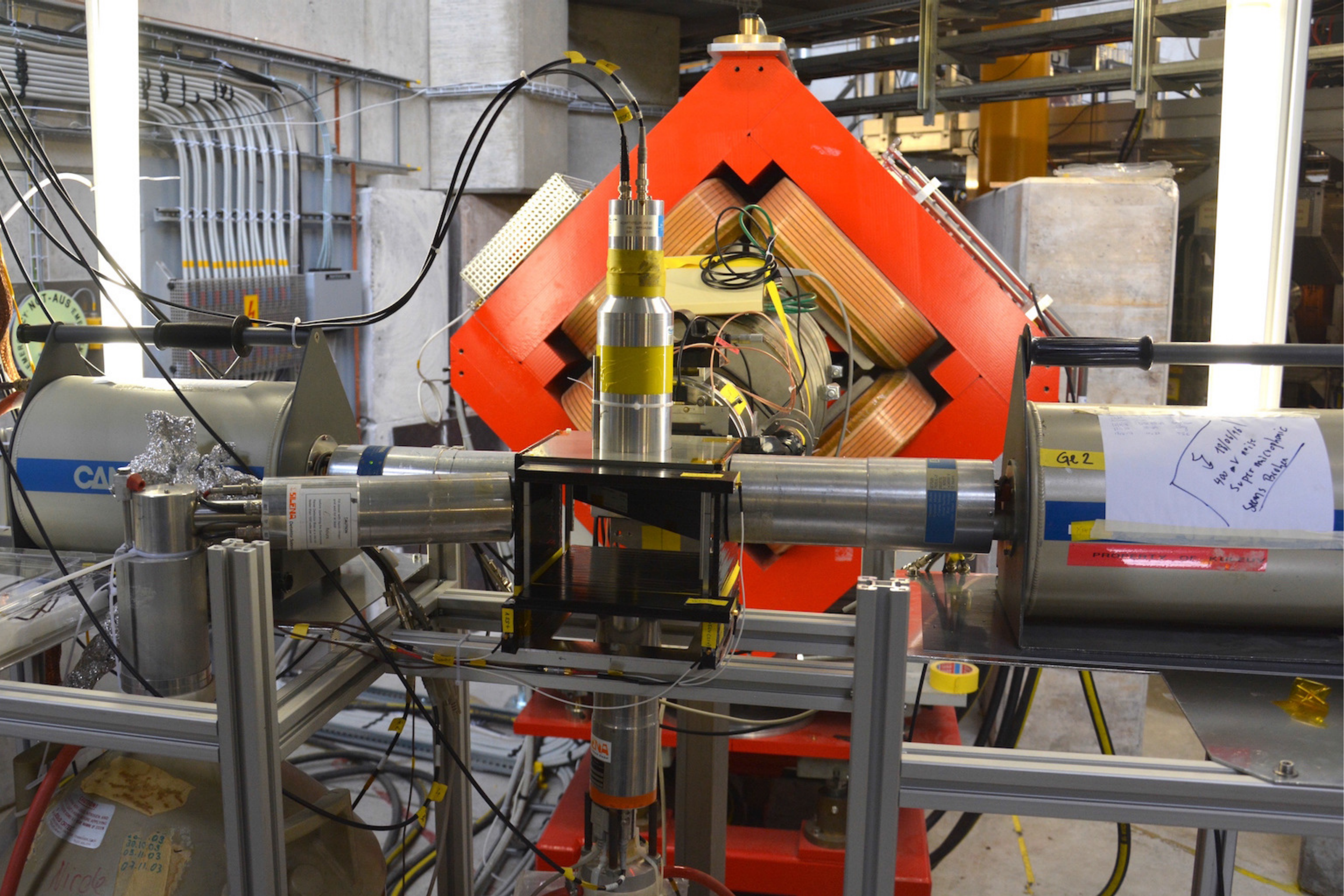}}
\caption{Detection setup and target holder mounted at the end of the $\pi$E1 area of the HIPA facility 
at the Paul Scherrer Institut. The last quadrupole of the muon beam line is visible at the back of the detection setup. 
The four plastic scintillator counters are mounted in a box-like structure surrounding the target holder. Four HPGe detectors 
and a LaBr$_3$ scintillator (top) were used to detect the muonic X rays. 
\label{fig02}} 
\end{figure}

The muonic X rays following the muonic cascade were detected by two single crystal 
high-purity germanium (HPGe) coaxial detectors 
with relative efficiency of 20\% and 75\% placed in close vicinity to the 
target at 90$^\circ$ (GeR) and -90$^\circ$ (GeL), respectively, 
with respect to the direction of the incoming beam. Fig.~\ref{fig02} shows the detector arrangement. Two more 
HPGe detectors and a LaBr$_3$ scintillator were also operated but they are not used for the analysis 
presented here. The typical energy resolution for 1.3~MeV $\gamma$
radiation was 2.1~keV and 2.9~keV (FWHM), for the 20\% and 75\% detector, respectively.
The absolute photo-peak efficiency for the
359.8~keV line, the most intense transition in the $5g_{9/2} \rightarrow 4f_{7/2}$ hfs
observed in $^{185}$Re, was $\sim$0.2\% and $\sim$0.5\% for GeR and GeL, respectively, 
for the given geometry. The efficiency
calibration was performed using standard sources of $^{137}$Cs, $^{60}$Co, $^{88}$Y and $^{152}$Eu. 
Typical single rates in the germanium detectors were 500 and 1000 counts per second, respectively.\\
 Finally four plastic scintillator counters 5~mm thick and 18 $ \times$ 18~cm$^2$ large were placed around 
the target in a box-like structure. The signals from these plastic scintillator counters were used in anti-coincidence with 
the germanium detectors signals and allowed removal of  background events in the X-ray spectra mainly produced 
by the electrons emitted in the muon decay.  

The readout system was based on the STRUCK SIS3316 digitiser and the MIDAS data acquisition system~\cite{MIDAS}. 
This is a VME module providing 16 spectroscopic channels with a 250~MHz 14 bits sampling ADC each. The signals from each of 
the detector preamplifiers are passed directly to the SIS3316 modules. The smaller signals from GeL were routed through a fast 
amplifier in order to match better the dynamic range of the digitiser. The
filtering is performed digitally using algorithms implemented on field
programmable gate arrays (FPGA) on the SIS3316 board, a fast filter being used for triggering, timing and pile-up rejection and a slow filter
for energy determination. With a data acquisition running in trigger-less mode, time and energy were recorded for all detector signals above a certain threshold.
Additionally, for the germaniums and LaBr$_3$ scintillator, the traces were also read out. Prompt
and delayed muonic X-ray spectra of the germanium detectors were built by imposing conditions in the time difference
between the germanium detector and the muon counter. In a similar way the anti-coincidence conditions of the germanium signals 
with the signal of the other scintillator counters were built and applied to reduce background in the X-ray energy spectra.    

%%%%%%%%%%%%%%%%%%%%%%%%%%%%%%%%%%%%%%%%%%%%%%%%%%%%%%%%%%%%%%%%%%%%%
%%%%%%%%%%%%%%%%%%%%%%%%%%%%%%%%%%%%%%%%%%%%%%
\subsection{\label{sec:calibration}Calibration}
The usual experimental sequence involved collecting data from a rhenium target for 4~hours (4 runs),
with two hours calibration runs with the lead target directly preceding and following each group of 4 runs with rhenium. 
The main purpose of these calibration runs was to verify that there
had been no substantial gain shift during the target runs which might cause loss of energy resolution. \\
Line shifts due to electronic instability were checked using a $^{60}$Co radioactive source and the 2614.5~keV $\gamma$ ray 
in the natural radioactive background. The source was placed near the target and its $\gamma$-rays appeared in the muonic X-ray
spectrum. In order to sum all individual calibration runs, each run must be corrected for any relative gain shift and shift 
of the base line. This was done by first locating the centroids of the 1332.5 and 2614.5~keV $\gamma$ rays 
appearing in all runs.
By comparing the centroids of the peaks from these runs with a preselected run, one can determine the gain shift and the shift of 
the zero offset. To ensure sufficient statistics the spectra were evaluated every two runs. 
After correcting gain shifts, the spectra of the different runs were summed. Typical gain shifts were in the order of 0.03\%.
In the energy calibration the well-established energies of the muonic X rays in 
$^{208}$Pb, $^{16}$O and $^{12}$C were used. 
Muonic X rays from oxygen and carbon were observed in the prompt energy spectra due to the accidental hit of the muon beam 
on materials surrounding the targets. 
%-------------------------------------------------------------------------------------------
\section{\label{sec:results}Results}
\begin{figure}[h]
\centering{
\includegraphics[scale=.47]{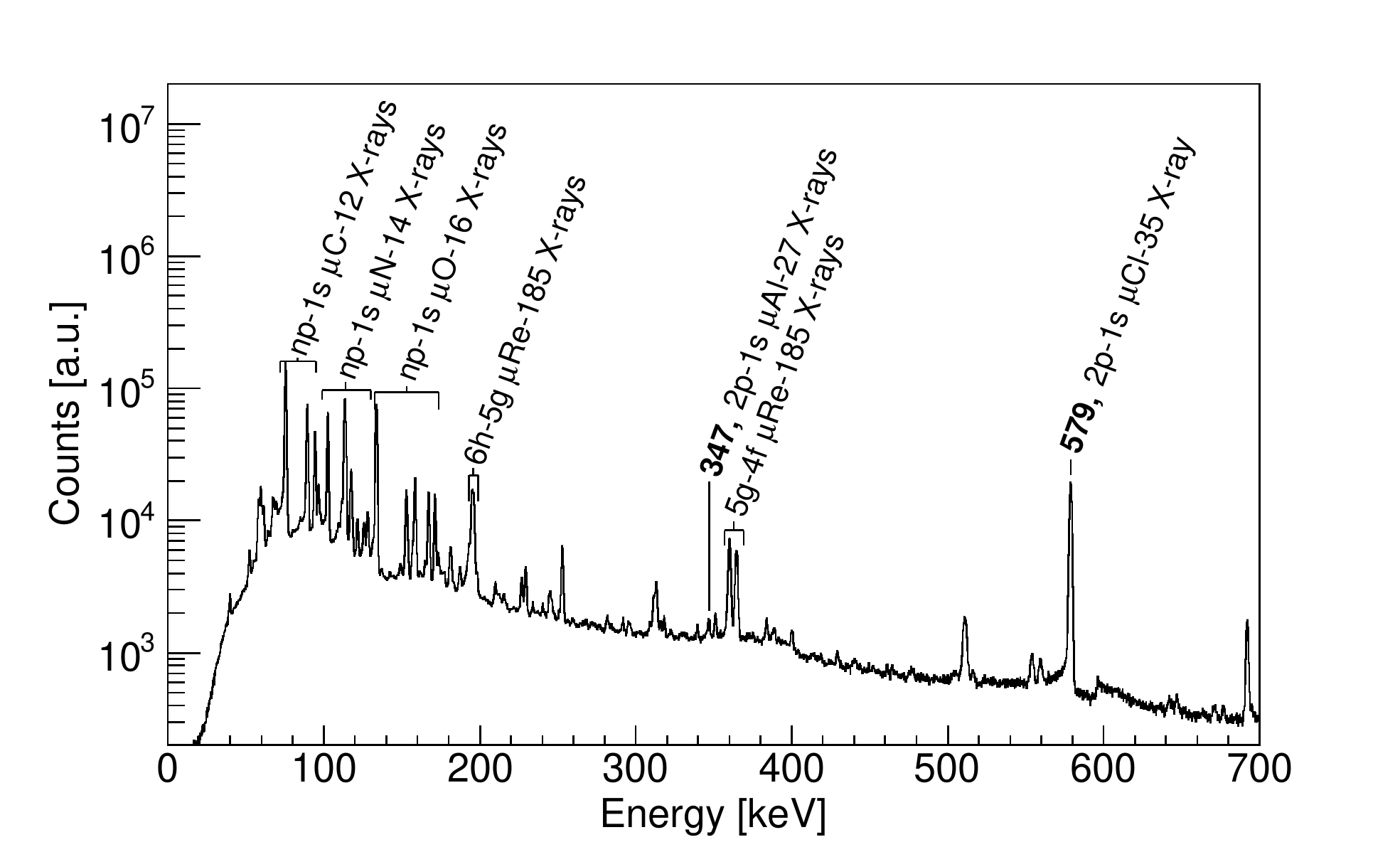}
\includegraphics[scale=.47]{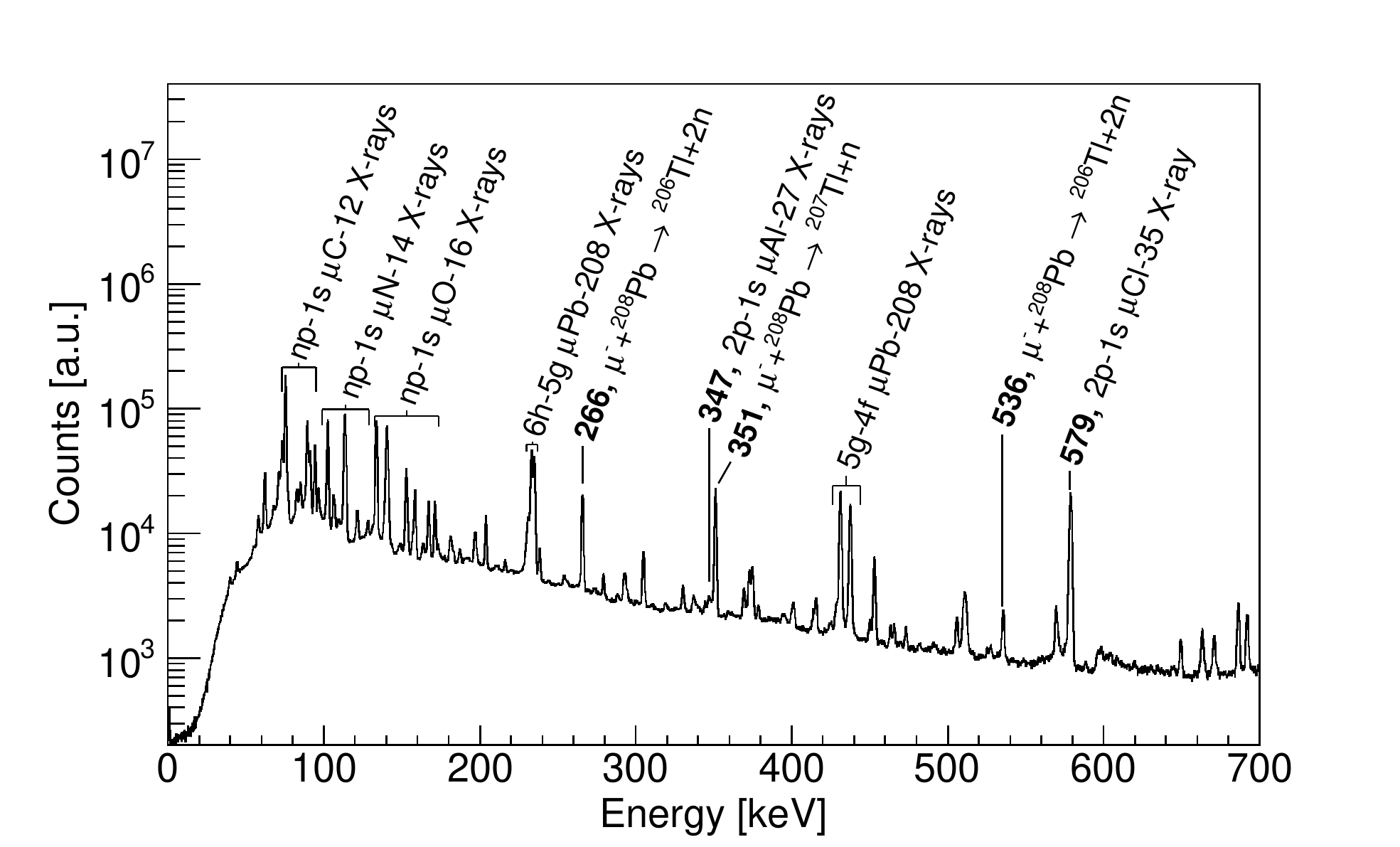}
}
\caption{The $\gamma$-ray energy spectra obtained with the $^{185}$Re target (top) and the 
$^{208}$Pb target (bottom) in GeR in prompt coincidence (0-400 ns) with the muon entrance counter. 
\label{fig03}} 
\end{figure}

\begin{figure}[h]
\centering{\includegraphics[scale=.47]{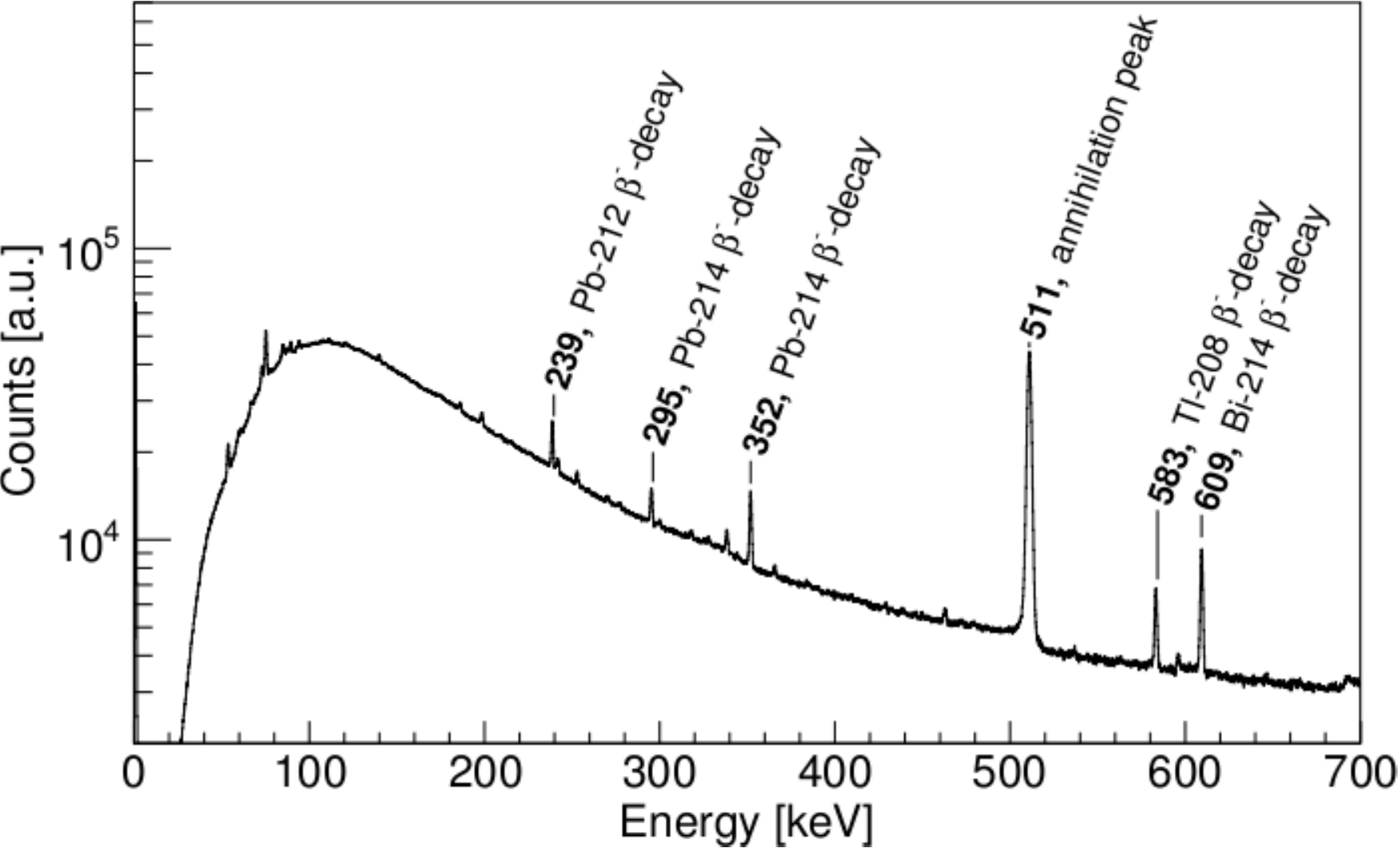}}
\caption{Background $\gamma$-ray energy spectrum obtained in GeR.
\label{fig04}} 
\end{figure}

By applying time conditions in the coincidence events between the Ge detectors and the entrance muon counter, 
it was possible to select prompt Ge events such as muonic X rays, where the muon stop and the subsequent atomic X rays 
are instantaneous within the time resolution of the detectors, and nuclear $\gamma$ rays resulting from the  muon-capture 
process which exhibits a characteristic lifetime of about 80 ns at Z$\sim$75 ~\cite{Suz87}.
Fig.~\ref{fig03} shows a portion of the $\gamma$-ray spectrum in the energy region of interest 
measured in the Ge detector positioned at 90$^\circ$ (GeR) with the  $^{185}$Re (top) and $^{208}$Pb (bottom) 
targets in prompt coincidence (400~ns) with the entrance muon counter. In addition Ge detector events not in coincidence with 
the entrance muon counter within 2~$\mu$s were selected to produce room background $\gamma$-ray spectra, 
as shown in Fig.~\ref{fig04}. In Fig.~\ref{fig03} the transitions belonging to muonic $^{208}$Pb and $^{185}$Re are indicated 
together with the muonic X rays of $^{35}$Cl,  $^{27}$Al, $^{16}$O, $^{14}$N and $^{12}$C. The assignment of $\gamma$ lines 
was based on previously known transitions~\cite{Kes75, Back67}. 
Other strong lines in Fig.~\ref{fig03} and Fig.~\ref{fig04} come from the decay of nuclei produced in the muon capture reaction or 
from room background. One of the strongest lines in the spectrum is the 
511~keV $\gamma$ ray, originating mainly from the annihilation of the positrons produced in the 
electromagnetic cascade of the high-energy electron emitted in the muon decay.\\
Data for around 60~hours were collected with muons on the $^{208}$Pb target, 38~hours on the $^{185}$Re target 
and 59~hours on the $^{187}$Re target.

%%%%%%%%%%%%%%%%%%%%%%%%%%%%%%%%%%%%%%%%%%%%%%%%%%%%%%%%%%%%%%%%%%%%%
%%%%%%%%%%%%%%%%%%%%%%%%%%%%%%%%%%%%%%%%%%%%%%

%%%%%%%%%%%%%%%%%%%%%%%%%%%%%%%%%%%%%%%%%%%%%%%%%%%%%%%%%%%%%%%%%%%%%
%%%%%%%%%%%%%%%%%%%%%%%%%%%%%%%%%%%%%%%%%%%%%%

%-------------------------------------------------------------------------------------------
\subsection{\label{sec:peakfit}Line shape}
Since the hyperfine splitting is the result of the convolution of many transitions, particular care has to be taken in describing 
the experimental line shape of each transition.
The mathematical form of the line shape should represent the response of the Ge detector plus a background term. \\ 
In this respect, the model used consists of a Gaussian peak $g(E)$, a step-like shelf $s(E)$, 
and a Hypermet function $t(E)$~\cite{Cam97, Sal16}. The latter is added to account 
for a possible tail, which decays exponentially below the peak's centroid and is produced by incomplete charge collection
and ballistic deficits. The model function was fitted to the shape of the peak by
using RooFit~\cite{Roo03}. RooFit implements its data models in terms of probability density functions (PDFs),
 which are by definition unit normalised. The model function describing the number of counts in the peak at energy
$x_0$ may be written as 

\begin{equation}
\label{eq:lineshape}
\begin{aligned}[b]
f(E) & = N_{\rm signal}  \\
& \times [f_{\rm gauss} \cdot g(E) + f_{\rm tail} \cdot t(E) + s(E)] \\
& + B
\end{aligned}
\end{equation}

where
\begin{equation*}
\begin{aligned}[b]
& g(E) = \frac{1}{\sqrt{2\pi} \sigma} \cdot \exp\left(-\frac{(E-x_0)^2}{2\sigma^2}\right)  \\
& t(E) = \frac{1}{2\beta} \cdot \exp\left(\frac{E-x_0}{\beta}+\frac{\sigma^2}{2\beta^2}\right) \cdot \verb+erfc+\left( \frac{E-x_0}{\sqrt{2}\sigma}+
\frac{\sigma}{\sqrt{2}\beta} \right) \\
& s(E) = \frac{A}{2} \cdot \verb+erfc+\left(\frac{E-x_0}{\sqrt{2}\sigma}\right)
\end{aligned}
\end{equation*}

\begin{figure}[h]
\centering{
\includegraphics[scale=.42]{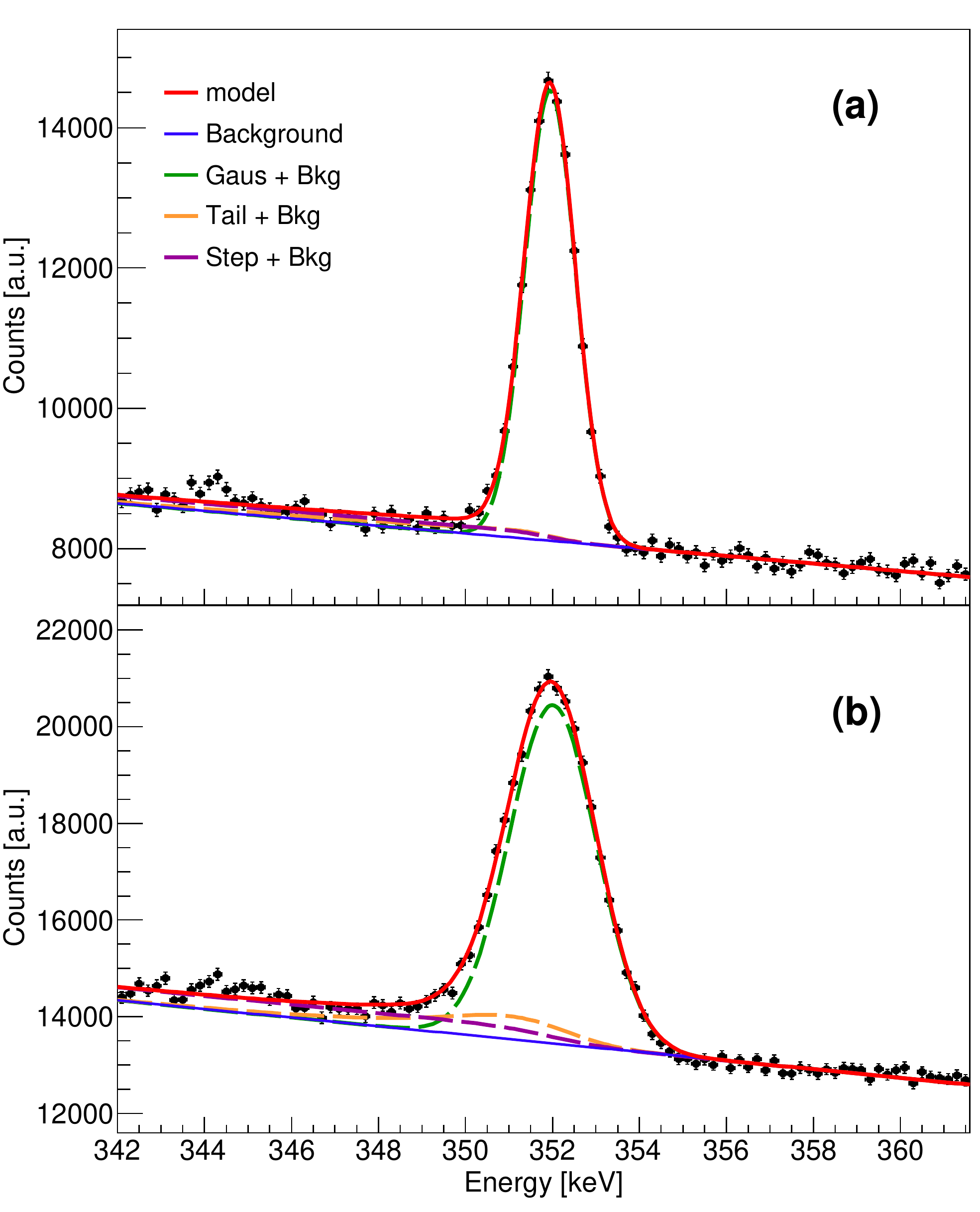}
}
\caption{[color online] The 351.9 keV $\gamma$-ray line of $^{214}$Pb decay (natural background) observed in the spectrum of the GeR (a) detector
and the GeL (b) detector and the different components of the peak shape model. 
\label{fig05}} 
\end{figure}

In these formulae $x_0$ is the mean of the Gaussian, $\sigma$ the Gaussian width and $\beta$ the slope of the exponential
tail. $f_{\rm gauss}$ denotes the fraction of the line shape having the Gaussian form and $f_{\rm tail}$ = (1- $f_{\rm gauss}$) 
the fraction having the exponential tail. The parameter A denotes the amplitude of the step which 
is proportional to the number of events in the signal. The parameter B is introduced to describe a constant background. 
This description was valid for most of the transitions except for the few cases where a linear function provided a better 
description. \\
The variables $\sigma$, $x_0$, $\beta$, the number of events in the  signal $N_{\rm signal}$, $f_{\rm gauss}$, 
and the two amplitudes A and B are free parameters of the model.
Since the response of the germanium detector is energy dependent, a consistent set of parameters 
($\sigma$, $\beta$, $f_{\rm gauss}$, A)
describing the experimental line shape  was obtained by fitting four nuclear transition lines which lie close in energy to the
muonic transitions of interest. These are the 265.8~keV transition from muon capture in $^{208}$Pb observed in the prompt 
spectrum with the $^{208}$Pb target and the 351.9~keV, 583.2~keV and 609.3~keV transitions observed in the room background. 
The natural line width of these lines is assumed to be negligible compared to the experimental resolution. 
The four transitions were fitted simultaneously with the Gaussian width $\sigma$ expressed as a linear function of the 
peak position $\sigma(E) = a_{\sigma} E + b_{\sigma}$~\cite{Kno10}.
The set of line-shape parameters obtained with this procedure are reported in Table~\ref{tab:table2}. 
Fig.~\ref{fig05} shows the 351.9~keV transition observed in the spectrum of the GeR and
the GeL detectors together with the fit function described in Eq.~(\ref{eq:lineshape}). 
Similar fits were obtained for the 265.8, 583.2 and 609.3~keV transitions. \\
It is important to note that by determining the line shape from the set of data collected
with beam on target, we ensure the appropriate representation of  the detector response in the presence of beam. 
In previous analyses~\cite{Kon81, Dixit1971} the line shape used to determine the position of the muonic 
X rays was the same as the one used for the calibration source lines collected in dedicated runs. 
With this procedure one relies on the strong assumption that the
detector response stayed unchanged between the X-ray runs and the calibration runs.

\begin{table}[h]%The best place to locate the table environment is directly after its first reference in text
\caption{\label{tab:table2}
Set of parameters resulting from the simultaneous fit of four $\gamma$-ray transitions (see text).}
%\begin{indented}
\begin{ruledtabular}
\begin{tabular}{l l l }
 Fit parameter & GeL  & GeR  \\
\colrule
$a_{\sigma}$  & 0.00024(1) & 0.00034(1)   \\
$b_{\sigma}$ (keV)& 0.918(7) & 0.466(4)  \\
$\beta$ (keV) & 2.2(2) & 5.0(8)\\
$f_{\rm gauss}$ &  0.893(6) & 0.93(1)\\
A (1/keV) & 0.0137(7) & 0.010(1)\\
\end{tabular}
\end{ruledtabular}
\end{table}
 
The muonic X-ray peaks are broader than the calibration source lines or background lines due to the natural  width 
of the muonic energy states. Since the intrinsic X-ray line shape is Lorentzian, the muonic X rays were fitted using 
the experimental line shape of Eq.~(\ref{eq:lineshape}) where the Gaussian component is modified into a 
Gaussian-convoluted Lorentzian (resulting in a Voigt profile) with calculated transition widths.
The typical natural line widths are $\sim$80~eV for the $5g_{9/2} \rightarrow 4f_{7/2}$, 
$5g_{7/2} \rightarrow 4f_{5/2}$ and $5g_{7/2} \rightarrow 4f_{7/2}$ transitions and $\sim$150~eV for the 
$5f_{7/2} \rightarrow 4d_{5/2}$ and $5f_{5/2} \rightarrow 4d_{5/2}$ transitions.\\

%%%%%%%%%%%%%%%%%%%%%%%%%%%%%%%%%%%%%%%%%%%%%%%%%%%%%%%%%%%%%%%%%%%%%
%%%%%%%%%%%%%%%%%%%%%%%%%%%%%%%%%%%%%%%%%%%%%%%
%-----------------------------------------------------------------------------------------------------------------------------------------------
\subsection{\label{sec:hfsanalysis}
The hyperfine splitting in $^{185,187}$Re}
\begin{figure}[h]
\centering{\includegraphics[scale=0.47]{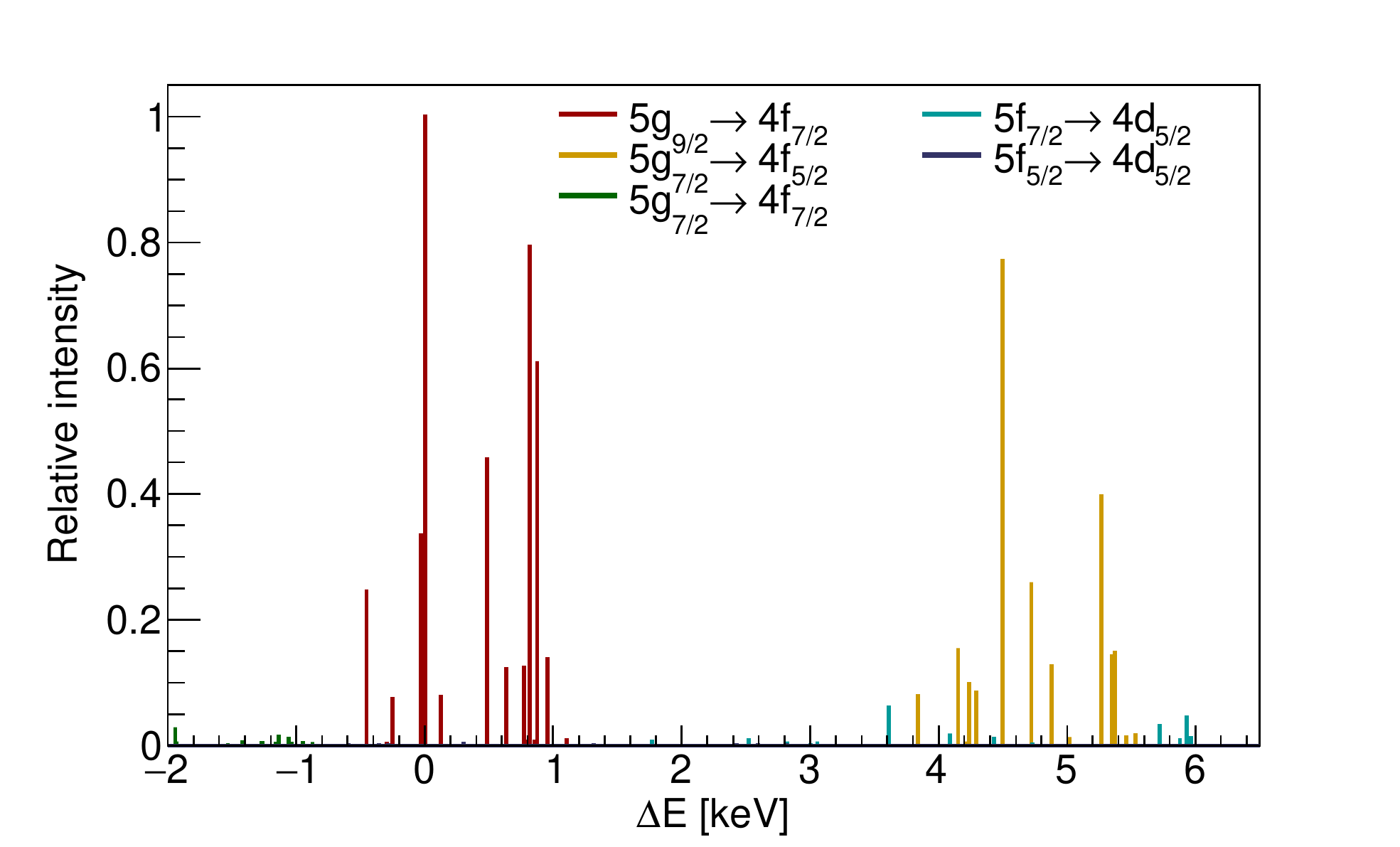}}
\caption{[color online] Relative positions and intensities of the transitions between the 5$g$ and 4$f$ multiplets  and 
5$f$ and 4$d$ (see also Table~\ref{tab:table3}) in $^{185}$Re considered in the present analysis. 
The energy displacements and heights are calculated for $Q$=2.21 barn. The height of the lines is proportional to the
intensity of the transitions.  
\label{fig06}} 
\end{figure}

The analysis of the hyperfine splitting was performed in higher muonic levels n=5 and n=4.
The $5g_{9/2} \rightarrow 4f_{7/2}$ and $5g_{7/2} \rightarrow 4f_{5/2}$ hfs 
complexes appear as two bumps located at around 360 keV (see Fig.~\ref{fig03}) and they have been analysed together. 
The $5g$ and the $4f$ levels of $^{185,187}$Re are sixfold split as in this case $I=5/2$ and 
\textit{l} ~=~5, 4. Taking the selection rules into account for transitions within both hf complexes, the resulting 
X-ray pattern consists of thirty members.\\ In the analysis of the hfs spectrum the correction for
the presence of the weaker $5f_{7/2} \rightarrow 4d_{5/2}$, $5g_{7/2} \rightarrow 4f_{7/2}$ and 
$5f_{5/2} \rightarrow 4d_{5/2}$ multiplets has to be taken into account, as they coincide in energy.
The hfs spectra analysed consisted therefore of 76 lines originating from five multiplets  which
were fitted using for each line the empirical line shape described by Eq.(\ref{eq:lineshape}) 
corrected for the radiative width. The background constant $B$ was common for all the lines.\\ 
The intensity and energy position of the individual members of the hf multiplets relative to the 
most intense transition for each multiplet were calculated using the
formalism described in Section~\ref{sec:hfs} and in Section~\ref{sec:fit}. The values are given in Table~\ref{tab:table3} 
and in Table~\ref{tab:table3a} for $^{185}$Re and $^{187}$Re, respectively.
The multiplets were then correlated in energy to the $F = 7 \rightarrow 6$ transition 
in $5g_{9/2} \rightarrow 4f_{7/2}$ using the values given in Table~\ref{tab:table4}. \\
It should be noted that the energy splittings of the hyperfine transitions in the $5f_{7/2} \rightarrow 4d_{5/2}$
multiplet are around a factor three larger than the values given in Ref.~\cite{Kon81} whereas the calculations for the
other multiplets agree. The difference can be due to a mistake in reporting the values.
The intensity of the three multiplets originating from the $5g$ state has been correlated to the intensity 
of the most intense $F = 7 \rightarrow 6$ hyperfine transition in $5g_{9/2} \rightarrow 4f_{7/2}$ 
assuming the hypothesis that the states within a \textit{l} multiplet are statistically populated; similarly 
the intensity of two multiplets originating from the $5f$ has been correlated to the intensity 
of the most intense $F = 6 \rightarrow 5$ hyperfine transition in $5f_{7/2} \rightarrow 4d_{5/2}$. 
The relative intensity of the lines within a \textit{l} multiplet does not depend on the initial distribution of the cascade and 
therefore they were kept fixed in the fitting procedure. On the other hand, no assumption can be made on the 
relative population of the $5g_{9/2}$ and $5f_{7/2}$ states as it depends on the details of the 
atomic cascade of muons which are still rather uncertain, particularly as to the exact beginning of the cascade.\\
Following this procedure, the description of the hfs could be reduced to five parameters which are
the energy of the $F = 7 \rightarrow 6$ hyperfine transition in $5g_{9/2} \rightarrow 4f_{7/2}$, 
the quadrupole moment, the two intensities of the $5g_{9/2} \rightarrow 4f_{7/2}$ and 
$5f_{7/2} \rightarrow 4d_{5/2}$ transitions and the number of background events. 
They were used as free parameters and varied until the best fit to the spectra was found.
Fig.~\ref{fig06} shows the theoretical prediction of the hfs of the five multiplets considered in 
the present analysis calculated for $Q$ = 2.21 barn. In the figure the intensity ratio  $5f_{7/2} \rightarrow 4d_{5/2}$
over $5g_{9/2} \rightarrow 4f_{7/2}$ is set equal to 0.06 as obtained from a cascade calculation~\cite{Aky78} 
with initial statistical distribution at N = 20 and width of the K-shell refilling process of 25 eV. Different initial
conditions of the cascade calculations give a range of values between 0.06 and 0.08.
 
%%%%%%%%%%%%%%%%%%%%%%%%%%%%%%%%%%%%%%%%%%%%%%%%%%%%%%%%%%%%%%%
%%%%%%%%%%%%%%%%%%%%%%%%%%%%%%%%%%%%%%%%%%%%%%
\begin{turnpage}
\begingroup
\squeezetable
\begin{table*}
\caption{\label{tab:table3}Theoretical parameters describing the dependence upon the quadrupole moment $Q$~(barn) of the relative intensity I 
and energy shifts $\Delta$E (eV) of the individual members of the five hf complexes fitted in $^{185}$Re. 
For each transition the values of I$_{0,1,2}$ and $\Delta$E$ _{0,1,2}$ calculated as described in 
Section~\ref{sec:fit} are expressed relative to the most intense transition of each multiplet and reported respectively in subsequent rows.}

\begin{ruledtabular}
\begin{tabular}{cdddddddddd} 
 &\multicolumn{2}{c}{ $5g_{9/2} \rightarrow 4f_{7/2}$ } &\multicolumn{2}{c}{ $5g_{7/2} \rightarrow 4f_{5/2}$ }
 &\multicolumn{2}{c}{ $5g_{7/2} \rightarrow 4f_{7/2}$ } &\multicolumn{2}{c}{ $5f_{7/2} \rightarrow 4d_{5/2}$ }  
 &\multicolumn{2}{c}{ $5f_{5/2} \rightarrow 4d_{5/2}$ } \\

F$_f \rightarrow $F$_i$ & I_{0,1,2} & \Delta E_{0,1,2}  &  I_{0,1,2} &  \Delta E _{0,1,2}& I_{0,1,2}  &  \Delta E_{0,1,2} 
&I_{0,1,2}  &  \Delta E_{0,1,2} &I_{0,1,2}  &  \Delta E_{0,1,2}  \\

\hline
 
$7 \rightarrow 6$  &100.0  & 0        &       &         &       &         &       &         &        &          \\
                   &0      &0        &       &         &       &         &       &         &        &          \\
                   &0      &0        &       &         &       &         &       &         &        &          \\
$6 \rightarrow 6$  &8.045  &-4.371     &       &         &100.0  &0        &       &         &        &          \\  
                             &-0.115 &-112.842 &       &         &0         &0        &       &         &        &          \\   
                             &-0.011 &0.435      &       &         &0         &0        &       &         &        &          \\  
$6 \rightarrow 5$  &78.602 &8.166    &100.0  &0        &10.899 &12.537   &100.0  &0        &        &          \\
                   &0.383  &374.540  &0      &0        &1.664  &487.383  &0      &0        &        &          \\
                   &-0.008 &-3.999   &0      &0        &0.722  &-4.434   &0      &0        &        &          \\
$5 \rightarrow 6$  &0.311  &-8.601   &       &         &10.234 &-6.088   &       &         &        &          \\
                   &-0.002 &-130.457 &       &         &-0.401 &-116.652 &       &         &        &          \\
                   &-0.000 &0.008    &       &         &-0.010 &0.427    &       &         &        &          \\
$5 \rightarrow 5$  &12.375 &3.936    &12.062 &-6.088   &68.235 &6.449    &12.106 &-6.395   &100.0   &0         \\
                   &0.036  &356.926  &0.218  &-116.652 &-8.661 &370.730  &-0.307 &-252.529 &0       &0         \\
                   &-0.012 &-4.426   &0.022  &0.427    &-0.117 &-4.007   &-0.041 &2.357    &0       &0         \\
$5 \rightarrow 4$  &60.644 &20.395   &72.591 &15.161   &15.636 &22.909   &72.304 &-7.636   &17.965  &-1.242    \\
                   &0.214  &385.392  &0.360  &397.303  &-1.370 &399.196  &1.215  &1142.401 &3.914   &1394.930  \\
                   &0.000  &-0.001   &0.036  &-4.013   &0.022  &0.417    &-0.106 &-39.216  &1.842   &-41.573   \\
$4 \rightarrow 5$  &0.669  &0.596    &0.730  &-11.348  &15.656 &1.190    &0.738  &-14.947  &16.736  &-10.873   \\
                   &0.012  &397.203  &0.024  &-124.867 &-1.686 &362.516  &-0.002 &-267.191 &-1.155  &-267.545  \\
                   &-0.001 &-4.335   &0.003  &0.337    &-0.009 &-4.097   &0.     &0.007    &-0.008  &2.362     \\
$4 \rightarrow 4$  &13.527 &17.055   &18.102 &9.902    &44.693 &17.649   &18.040 &-16.188  &56.816  &-12.115   \\
                   &0.109  &425.669  &0.088  &389.089  &-3.994 &390.982  &0.343  &1127.739 &-16.181 &1127.385  \\
                   &-0.001 &0.089    &0.008  &-4.104   &0.101  &0.327    &-0.106 &-41.565  &0.792   &-39.210   \\
$4 \rightarrow 3$  &45.818 &28.096   &50.423 &29.026   &16.754 &28.690   &50.338 &47.507   &24.306  &51.581    \\
                   &0.007  &210.027  &0.397  &339.460  &-3.201 &175.340  &0.381  &940.730  &-5.451  &940.376   \\
                   &-0.002 &-1.336   &0.021  &-2.673   &0.210  &-1.098   &-0.007 &-2.997   &0.382   &-0.643    \\
$3 \rightarrow 4$  &0.837  &14.338   &1.642  &5.738    &16.889 &13.486   &1.639  &-21.887  &24.769  &-21.950   \\
                   &0.019  &494.007  &0.009  &440.624  &-1.220 &442.517  &0.086  &1239.448 &-5.071  &1153.201  \\
                   &-0.000 &0.244    &0.001  &-4.259   &0.019  &0.172    &-0.016 &-40.816  &0.130   &-39.964   \\
$3 \rightarrow 3$  &11.900 &25.378   &19.212 &24.862   &28.837 &24.526   &19.267 &41.809   &25.862  &41.746    \\
                   &0.104  &278.364  &-0.046 &390.996  &0.243  &226.875  &0.330  &1052.440 &-1.881  &966.192   \\
                   &-0.001 &-1.181   &0.012  &-2.828   &-0.120 &-1.253   &-0.002 &-2.248   &-0.002  &-1.396    \\
$3 \rightarrow 2$  &33.944 &33.736   &33.003 &39.126   &14.555 &32.884   &32.976 &70.357   &24.558  &70.294    \\
                   &-0.102 &-23.693  &0.257  &86.228   &-2.527 &-75.182  &-0.188 &231.453  &-7.174  &145.205   \\
                   &-0.002 &-2.376   &0.013  &-1.619   &0.149  &-2.448   &-0.019 &-21.133  & 0.644  &-20.280   \\
$2 \rightarrow 3$  &0.619  &23.393   &2.192  &21.687   &14.624 &21.351   &2.206  &37.497   &24.986  &34.411    \\
                   &0.013  &350.541  &-0.022 &463.687  &0.005  &299.566  &0.097  &1208.901 &-1.687  &1124.796   \\
                   &-0.000 &-1.215   &0.002  &-2.794   &-0.054 &-1.219   &-0.001 &-1.619   &-0.017  &-2.037    \\
$2 \rightarrow 2$  &7.725  &31.751   &16.477 &35.951   &19.289 &29.709   &16.510 &66.045   &8.988   &62.959    \\
                   &0.050  &48.484   &-0.097 &158.919  &1.893  &-2.491   &0.170  &387.914  &3.084   &303.809   \\
                   &-0.001 &-2.410   &0.009  &-1.584   &-0.166 &-2.414   &-0.009 &-20.504  &-0.278  &-20.921   \\
$2 \rightarrow 1$  &25.000 &36.560   &19.797 &46.222   &9.296  &34.518   &19.773 &68.493   &20.127  &65.408    \\
                   &-0.105 &-218.448 &0.094  &-172.923 &-0.887 &-269.423 &-0.275 &-470.144 &-3.215  &-554.249   \\
                   &-0.001 &-1.502   &0.009  &-2.478   &0.015  &-1.505   &-0.018 &-23.433  &0.079   &-23.851   \\
$1 \rightarrow 2$  &       &         &1.829  &33.776   &9.265  &27.534   &1.836  &63.578   &19.702  &57.663    \\
                   &       &         &-0.027 &223.648  &0.637  &62.238   &0.056  &526.161  &2.589   &476.520   \\
                   &       &         &0.001  &-1.365   &-0.066 &-2.195   &0.0    &-20.982  &-0.351  &-20.455   \\
$1 \rightarrow 1$  &       &         &10.989 &44.047   &16.650 &32.343   &10.995 &66.027   &1.602   &60.112    \\
                   &       &         &-0.078 &-108.194 &1.955  &-204.695 &0.009  &-331.897 &1.509   &-381.538  \\
                   &       &         &0.004  &-2.259   &-0.128 &-1.286   &-0.010 &-23.911  &0.043   &-23.384   \\
$1 \rightarrow 0$  &       &         &10.262 &49.292   &       &         &10.257 &73.024   &10.821  &67.109    \\
                   &       &         &-0.008 &-312.574 &       &         &-0.157 &-869.153 &0.943   &-918.794  \\
                   &       &         &0.005  &-3.979   &       &         &0.     &1.644    &-0.122  &2.170     \\
$0 \rightarrow 1$  &       &         &       &         &       &         &       &         &10.510  &57.410    \\
                   &       &         &       &         &       &         &       &         &3.105   &-275.170  \\
                   &       &         &       &         &       &         &       &         &-0.154  &-22.478   \\
                                                                                                                         
\end{tabular}
\end{ruledtabular}
\end{table*}
\endgroup
\end{turnpage}

%%%%%%%%%%%%%%%%%%%%%%%%%%%%%%%%%%%%%%%%%%%%%%%%%%%%%%%%%%%%%%%%%%%%%  

%%%%%%%%%%%%%%%%%%%%%%%%%%%%%%%%%%%%%%%%%%%%%%
\begin{turnpage}
\begingroup
\squeezetable
\begin{table*}
\caption{\label{tab:table3a}
Theoretical parameters describing the dependence upon the quadrupole moment $Q$~(barn) of the relative intensity I and energy shifts $\Delta$E (eV) 
of the individual members of the five hf complexes fitted in $^{187}$Re. For each transition the values of I$_{0,1,2}$ and $\Delta$E$ _{0,1,2}$ calculated as 
described in Section~\ref{sec:fit} are expressed relative to the most intense transition of each multiplet and reported respectively in subsequent rows.}
\begin{ruledtabular}
\begin{tabular}{cdddddddddd}
 &\multicolumn{2}{c}{ $5g_{9/2} \rightarrow 4f_{7/2}$ } &\multicolumn{2}{c}{ $5g_{7/2} \rightarrow 4f_{5/2}$ }
 &\multicolumn{2}{c}{ $5g_{7/2} \rightarrow 4f_{7/2}$ } &\multicolumn{2}{c}{ $5f_{7/2} \rightarrow 4d_{5/2}$ }  
 &\multicolumn{2}{c}{ $5f_{5/2} \rightarrow 4d_{5/2}$ } \\

F$_f \rightarrow $F$_i$ & I_{0,1,2} & \Delta E_{0,1,2}  &  I_{0,1,2} &  \Delta E _{0,1,2}& I_{0,1,2}  &  \Delta E_{0,1,2} 
&I_{0,1,2}  &  \Delta E_{0,1,2} &I_{0,1,2}  &  \Delta E_{0,1,2}  \\

\hline

$7 \rightarrow 6$&100.0  &         &       &         &       &         &       &         &        &          \\
                 &0      &0        &       &         &       &         &       &         &        &          \\
                 &0      &0        &       &         &       &         &       &         &        &          \\
$6 \rightarrow 6$&8.044  &-5.617   &       &         &100.0  &0        &       &         &        &          \\  
                 &-0.114 &-111.702  &       &         &0      &0        &       &         &        &          \\   
                 &-0.011 &0.163   &       &         &0      &0        &       &         &        &          \\  
$6 \rightarrow 5$&78.607 &8.349    &100.0  &0        &10.737 &13.966   &100.0  &0        &        &          \\
                 &0.379  &374.490  &0      &0        &1.815  &486.192  &0      &0        &        &          \\
                 &-0.007 &-3.988   &0      &0        &0.686  &-4.151   &0      &0        &        &          \\
$5 \rightarrow 6$&0.311 &-10.053   &       &         &10.229 &-7.368   &       &         &        &          \\
                 &-0.002 &-129.157 &       &         &-0.396 &-115.497 &       &         &        &          \\
                 &-0.000 &-0.301    &       &         &-0.011 &0.151    &       &         &        &          \\
$5 \rightarrow 5$&12.377 &3.913    &12.060 &-7.368   &68.172 &6.598    &12.110 &-7.010   &100.0   &0         \\
                 &0.034  &357.035  &0.219  &-115.497 &-8.597 &370.695  &-0.311 &-252.030 &0       &0         \\
                 &-0.012 &-4.452   &0.022  &0.151    &-0.132 &-4.000   &-0.041 &2.239    &0       &0         \\
$5 \rightarrow 4$&60.644 &20.099   &72.585 &15.375   &15.622 &22.784   &72.335 &-1.431   &17.855  &5.579    \\
                 &0.214  &385.870  &0.366  &397.289  &-1.356 &399.531  &1.186  &1136.395 &4.016   &1388.425  \\
                 &0.000  &-0.116   &0.035  &-4.011   &0.019  &0.337    &-0.099 &-37.895  &1.818   &-40.134   \\
$4 \rightarrow 5$&0.669  &0.966    &0.730  &-12.772  &15.645 &1.194    &0.738  &-15.399  &16.732  &-11.561   \\
                 &0.012  &396.905  &0.024  &-123.626 &-1.674 &362.566  &-0.002 &-266.904 &-1.151  &-267.018  \\
                 &-0.001 &-4.265   &0.003  &0.041    &-0.012 &-4.110   &0.00     &-0.060    &-0.009  &2.238     \\
$4 \rightarrow 4$&13.528 &17.153   &18.100 &9.970    &44.663 &17.380   &18.062 &-9.820  &56.521  &-5.982   \\
                 &0.108  &425.741  &0.090  &389.160  &-3.960 &391.402  &0.323  &1121.521 &-15.900 &1121.406  \\
                 &-0.001 &0.071    &0.008  &-4.122   &0.092  &0.226    &-0.101 &-40.194  &0.726   &-37.896   \\
$4 \rightarrow 3$&45.817 &27.876   &50.419 &29.285   &16.779 &28.104   &50.337 &48.678   &24.330  &52.516    \\
                 &0.008  &210.497  &0.401  &339.549  &-3.222 &176.158  &0.381  &939.615  &-5.470  &939.500   \\
                 &-0.002 &-1.449   &0.020  &-2.695   &0.215  &-1.294   &-0.007 &-2.835   &0.387   &-0.537    \\
$3 \rightarrow 4$&0.837  &15.137   &1.642  &6.278    &16.880 &13.688   &1.642  &-15.415  &24.721  &-15.924   \\
                 &0.019  &493.385  &0.009  &440.205  &-1.209 &442.446  &0.084  &1233.082 &-5.024  &1147.222  \\
                 &-0.000 &0.392    &0.001  &-4.161   &0.017  &0.188    &-0.015 &-39.410  &0.119   &-38.650   \\
$3 \rightarrow 3$&11.901 &25.860   &19.210 &25.593   &28.806 &24.412   &19.268 &43.083   &25.858  &42.574    \\
                 &0.103  &278.141  &-0.045 &390.593  &0.276  &227.203  &0.329  &1051.176 &-1.875  &965.316   \\
                 &-0.001 &-1.129   &0.012  &-2.733   &-0.128 &-1.333   &-0.002 &-2.050   &-0.004  &-1.290    \\
$3 \rightarrow 2$&33.943 &33.737   &33.001 &39.600   &14.578 &32.288   &32.972 &69.846   &24.615  &69.336    \\
                 &-0.101 &-23.380  &0.260  &86.218   &-2.547 &-74.318  &-0.184 &232.325  &-7.222  &146.464   \\
                 &-0.002 &-2.451   &0.012  &-1.616   &0.154  &-2.655   &-0.020 &-21.357  & 0.655  &-20.597   \\
$2 \rightarrow 3$&0.619  &24.606   &2.192  &23.133   &14.616 &21.952   &2.206  &38.935   &24.982  &35.315    \\
                 &0.013  &349.601  &-0.022 &462.572  &0.015  &299.181  &0.096  &1207.441 &-1.680  &1123.770   \\
                 &-0.000 &-0.993   &0.002  &-2.529   &-0.056 &-1.129   &-0.001 &-1.375   &-0.019  &-1.897    \\
$2 \rightarrow 2$&7.725  &32.482   &16.475 &37.140   &19.277 &29.829   &16.509 &65.698   &9.017   &62.078    \\
                 &0.050  &48.080   &-0.096 &158.196  &1.909  &-2.339   &0.170  &388.590  &3.057   &304.919   \\
                 &-0.001 &-2.315   &0.009  &-1.412   &-0.170 &-2.451   &-0.009 &-20.681  &-0.272  &-21.203   \\
$2 \rightarrow 1$&24.999 &36.969   &19.796 &46.935   &9.305  &34.316   &19.772 &68.716   &20.090  &65.097    \\
                 &-0.104 &-218.487 &0.095  &-173.098 &-0.894 &-268.907 &-0.274 &-469.540 &-3.176  &-553.210   \\
                 &-0.001 &-1.492   &0.009  &-2.436   &0.016  &-1.628   &-0.019 &-23.521  &0.070   &-24.043   \\
$1 \rightarrow 2$&       &         &1.829  &35.625   &9.264  &28.313   &1.836  &63.322   &19.709  &56.966    \\
                 &       &         &-0.027 &222.276  &0.639  &61.741   &0.056  &526.719  &2.585   &477.406   \\
                 &       &         &0.001  &-1.039   &-0.067 &-2.077   &0.0    &-21.132  &-0.350  &-20.684   \\
$1 \rightarrow 1$&       &         &10.987 &45.420   &16.655 &32.800   &10.996 &66.341   &1.641   &59.985    \\
                 &       &         &-0.077 &-109.018 &1.954  &-204.827 &0.009  &-331.411 &1.473   &-380.724  \\
                 &       &         &0.004  &-2.063   &-0.127 &-1.255   &-0.010 &-23.972  &0.052   &-23.523   \\
$1 \rightarrow 0$&       &         &10.261 &50.343   &       &         &10.257 &73.031   &10.826  &66.675    \\
                 &       &         &-0.008 &-313.047 &       &         &-0.157 &-868.097 &0.940   &-917.410  \\
                 &       &         &0.005  &-3.865   &       &         &0.     &1.478    &-0.122  &1.926     \\
$0 \rightarrow 1$&       &         &       &         &       &         &       &         &10.558  &57.409    \\
                 &       &         &       &         &       &         &       &         &3.062   &-274.499  \\
                 &       &         &       &         &       &         &       &         &-0.144  &-22.583   \\                                                       
                 
\end{tabular}
\end{ruledtabular}
\end{table*}
\endgroup
\end{turnpage}

%%%%%%%%%%%%%%%%%%%%%%%%%%%%%%%%%%%%%%%%%%%%%%%%%%%%%%%%%%%%%%%%%%%%%
\begin{table*}
\caption{\label{tab:table4}
Theoretical parameters describing the dependence upon the quadrupole moment $Q$~(barn) of the energy $\Delta$E (eV) of the most intense 
hyperfine transitions in each of the five hf complexes fitted in $^{185,187}$Re. For each transition the values of the parameters $\Delta$E$ _{0,1,2}$, 
calculated as described in Section~\ref{sec:fit}, are reported, respectively, in subsequent rows. 
The relative intensities are taken from cascade calculations~\cite{Aky78}
with initial statistical distribution at N = 20 and a width of the K-shell refilling process of 25 eV.}

\begin{ruledtabular}
\begin{tabular}{ccccc}

  $5g_{9/2} \rightarrow 4f_{7/2}$  & $5g_{7/2} \rightarrow 4f_{5/2}$ & $5g_{7/2} \rightarrow 4f_{7/2}$
 & $5f_{7/2} \rightarrow 4d_{5/2}$  & $5f_{5/2} \rightarrow 4d_{5/2}$ \\

  $7 \rightarrow 6$              &$6 \rightarrow 5$              &$6 \rightarrow 6$
  &$6 \rightarrow 5$              &$5 \rightarrow 5$  \\
 I = 0.333 & I = 0.257 & I = 0.0095 & I = 0.020 & I = 0.001\\
\hline
\multicolumn{5}{c}{$^{185}$Re}\\
$\Delta E_{0,1,2}$  &$\Delta E_{0,1,2}$   &$\Delta E_{0,1,2}$  &$\Delta E_{0,1,2}$  &$\Delta E_{0,1,2}$ \\ 
360.214  &364.663 &358.280  &364.417  &361.141  \\
-0.174   &-0.160  &-0.178   &-0.440   &-0.448 \\
 0.0     &0.004   &-0.000   &-0.002   &-0.004\\
\hline
\multicolumn{5}{c}{$^{187}$Re}\\
$\Delta E_{0,1,2}$  &$\Delta E_{0,1,2}$   &$\Delta E_{0,1,2}$  &$\Delta E_{0,1,2}$ &$\Delta E_{0,1,2}$ \\ 
360.215  &364.663 &358.280  &364.412  &361.136 \\
-0.175   &-0.160  &-0.178   &-0.439   &-0.448 \\
0.0      &0.004   &-0.000   &-0.002   &-0.004 \\
\end{tabular}
\end{ruledtabular}
\end{table*}

\begin{figure}[h]
\centering{
\includegraphics[scale=.44]{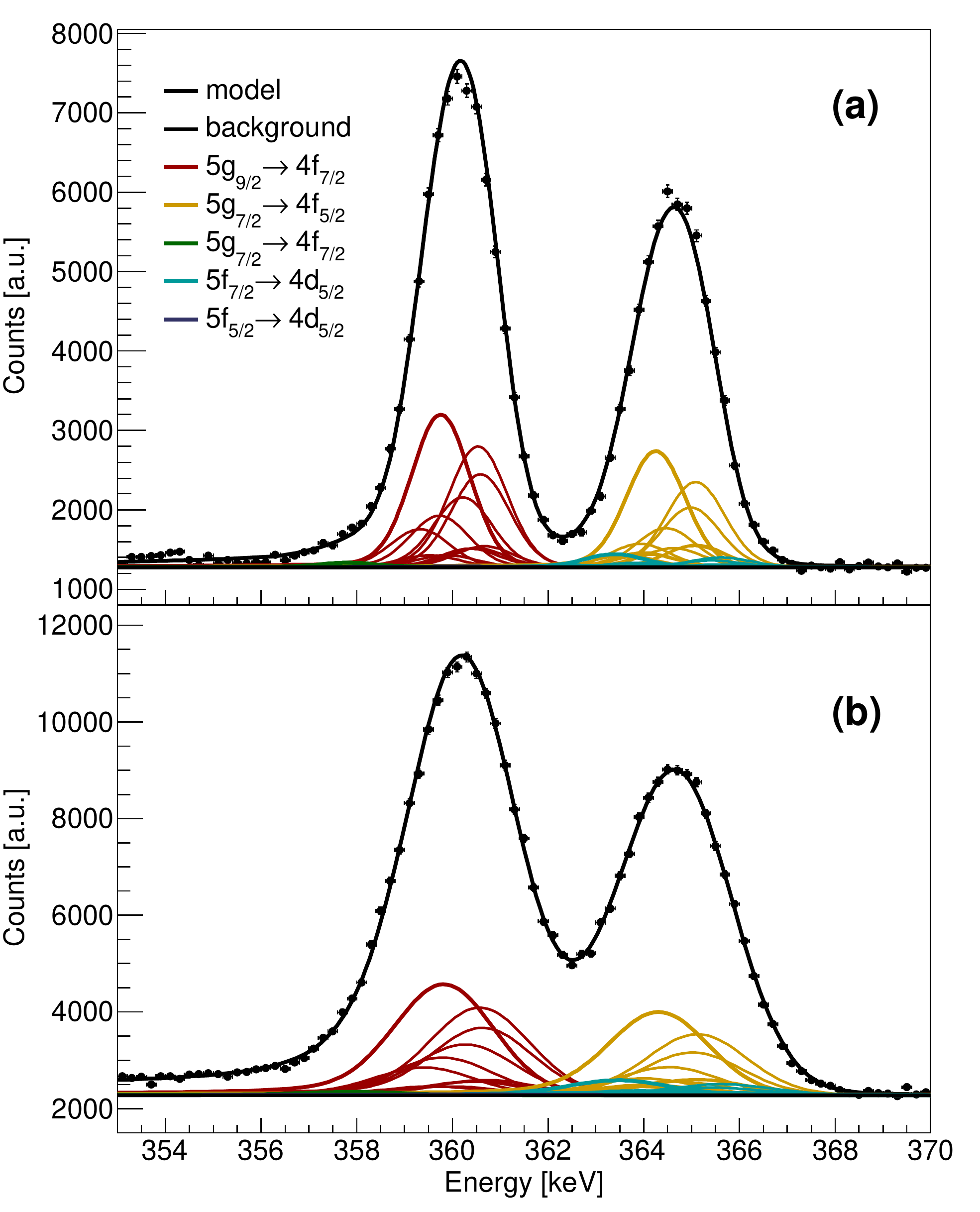}
}
\caption{[color online] Prompt $\gamma$-ray spectrum of $^{185}$Re obtained by GeR (a) and GeL (b)
showing the $5g \rightarrow 4f$ hf complex. The black line shows the best fit to the data. The lines predicted by the hfs 
formalism are shown below the spectra.
\label{fig07}} 
\end{figure}

\begin{figure}[h]
\centering{
\includegraphics[scale=.44]{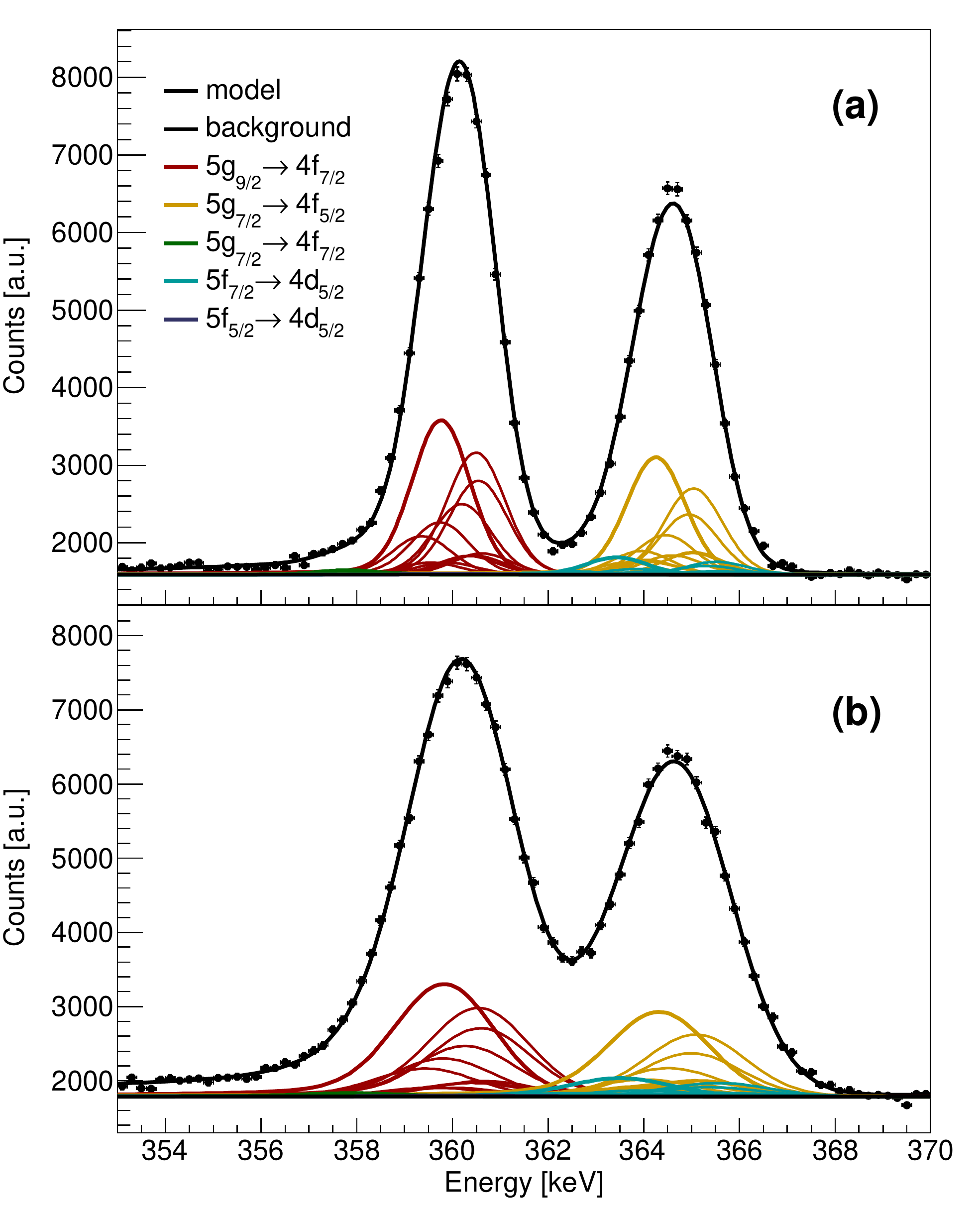}
}
\caption{[color online] Prompt $\gamma$-ray spectrum of $^{187}$Re obtained by GeR (a) and GeL (b)
showing the $5g \rightarrow 4f$ hfs complex. The black line shows the best fit to the data. The individual lines of the hfs 
formalism are shown below the spectra. 
\label{fig08}} 
\end{figure}

%%%%%%%%%%%%%%%%%%%%%%%%%%%%%%%%%%%%%%%%%%%%%%
The measured $5g\rightarrow4f$ spectrum of $^{185}$Re together with the result of the fit is shown 
in Fig.~\ref{fig07} for the two Ge detectors used. The fit for $^{187}$Re is shown in Fig.~\ref{fig08}. \\
Tables~\ref{tab:quadrupole1} and \ref{tab:quadrupole2} summarise the values of the fit parameters.
%%%%%%%%%%%%%%%%%%%%%%%%%%%%%%%%%%%%%%%%%%%%%%%%%%%%%%%%%%%%%%%%%%%
%%%%%%%%%%%%%%%%%%%%%%%%%%%%%%%%%%%%%%%%%%%%%%%%%%%%%%%%%%%%%%%%%%%%%
%%%%%%%%%%%%%%%%%%%%%%%%%%%%%%%%%%%%%%%%%%%%%%%%%%%%%%%%%%%%%%%%%%%%%
The isotopic impurity of the targets was included in the fitting procedure by having
 the hfs spectrum described by a double complex (one for each isotope) with one multiplet slightly shifted with 
respect to the other. The value of the shift is proportional to the ratio of the quadrupole moments
of the two isotopes which was taken from literature. In the case of the $^{187}$Re, the fit parameters 
were not affected by the inclusion of the small impurity of $^{185}$Re and it was therefore neglected in the final fit. \\
In addition to the five parameters mentioned above to describe the structure of the hfs, the step A of the line shape 
was also left as free parameter. This because it was not possible to reproduce the hfs with the value extracted 
from the line shape analysis. This effect might be due to the very different background between 
the delayed spectrum (where the line shape analysis has been performed) and that of the prompt spectrum.  \\
Reasonably good fits were obtained with $\chi^2$ per degree of freedom of 2.5 and 1.5  for GeR and GeL
in $^{185}$Re and 1.8 and 1.1 in $^{187}$Re. 
The energy of the $F = 7 \rightarrow 6$ hyperfine transition 
in $5g_{9/2} \rightarrow 4f_{7/2}$ multiplet obtained from the fit is 359.9(1) keV for GeL and 
359.8(1) keV for GeR. These values are the same for the two rhenium isotopes and show that, 
with the procedure described in Section~\ref{sec:calibration}, 
a very good energy calibration with precision at the level of 100 eV can be achieved.\\
The intensities of the $5f_{7/2} \rightarrow 4d_{5/2}$ transition relative to the 
$5g_{9/2} \rightarrow 4f_{7/2}$ obtained from the fits are higher compared to the value of 6\% 
obtained from the calculation of the muonic cascade. 
Such discrepancies are not surprising given the approximation of the cascade calculations 
and were observed in similar analyses~\cite{Dey1975,Dey1979}. 
Moreover, possible resonance effects between nuclear and muonic states could modify significantly
the muonic cascade leading to anomalous intensity ratios. Such resonance effects are more likely to 
occur in very deformed nuclei due to the dense nuclear excitation spectrum.
On the other hand, in the same isotope, the relative intensities deduced from the two detectors 
differs up to 50\%.
This inconsistency in the fitted relative intensities 
has been taken into account by adding a systematic error to the 
extracted quadrupole moment (see Section~\ref{sec:discussion}). 

%%%%%%%%%%%%%%%%%%%%%%%%%%%%%%%%%%%%%%%%%%%%%%
\begin{table*}
\caption{\label{tab:quadrupole1}Spectroscopic quadrupole moments $Q$, intensity RI of the $5f_{7/2} \rightarrow 4d_{5/2}$ transition 
relative to the $5g_{9/2} \rightarrow 4f_{7/2}$, step A and centroid energy E$_{7\rightarrow 6}$ obtained from the fit of the hfs of the 
$5g \rightarrow 4f$ transition in muonic $^{185}$Re. The influence on the extracted quadrupole moment of weak transitions is analysed.}
\begin{ruledtabular}
\begin{tabular}{ccccccccccc}
Fit & \multicolumn{5}{c}{GeR}&\multicolumn{5}{c}{GeL}\\

 & $Q$ (barn)    &RI &E$_{7 \rightarrow 6}$ (keV)  &A (1/keV)   & $\chi_{red}^2$  & $Q$ (barn)  &RI &E$_{7 \rightarrow 6}$ (keV) &A (1/keV)     & $\chi_{red}^2$   \\
\hline
Full \footnotemark[1]                           & 2.11(2) & 0.090(8)   & 359.8(1) &$\leq$10$^{-9}$   & 2.39    &2.04(5) & 0.139(7) &359.9(1) &0.0051(4)  &1.51  \\  
no $5f_{5/2} \rightarrow 4d_{5/2}$     & 2.11(2) & 0.090(8)    & 359.8(1) &$\leq$10$^{-9}$    & 2.34    &2.06(4) & 0.137(7) &359.9(1) &0.0051(4)  &1.50  \\  
no $5g_{7/2} \rightarrow 4f_{7/2}$     & 2.18(3) & 0.074(9)    & 359.7(1) &0.0009(6)  & 3.85   &2.17(4) & 0.122(7) &359.8(1) &0.0057(4)  &2.28  \\
no weak transitions \footnotemark[2]  & 2.18(3) & 0.076(9)   & 359.7(1) &0.0011(5)  & 3.80  &2.19(4) & 0.121(7) &359.8(1) &0.0058(4)  &2.22  \\
fix $5g/5f$ population\footnotemark[3]   & 2.03(2)& 0.139         &359.8(1)  &0.0014(4)  &2.85    &2.12(4) & 0.090      &359.8(1) &0.0037(3) & 2.06  \\
 
\end{tabular}
\end{ruledtabular}
\footnotetext[1]{Full fit as described in Section~\ref{sec:hfsanalysis}.}
\footnotetext[2]{The $5f_{5/2} \rightarrow 4d_{5/2}$ and $5g_{7/2} \rightarrow 4f_{7/2}$ multiplets are removed from the fit.}
\footnotetext[3]{Relative intensity fixed to 0.14 for GeR and 0.09 for GeL.}

\end{table*}
%%%%%%%%%%%%%%%%%%%%%%%%%%%%%%%%%%%%%%%%%%%%%%
\begin{table*}
\caption{\label{tab:quadrupole2}Spectroscopic quadrupole moments $Q$, intensity RI of the $5f_{7/2} \rightarrow 4d_{5/2}$ transition relative to the 
$5g_{9/2} \rightarrow 4f_{7/2}$ , step A and centroid energy E$_{7\rightarrow 6}$ obtained from the fit of the hfs of the $5g \rightarrow 4f$ transition 
in muonic $^{187}$Re. The influence on the extracted quadrupole moment of weak transitions is analysed.}
\begin{ruledtabular}
\begin{tabular}{ccccccccccc}
Fit & \multicolumn{5}{c}{GeR}&\multicolumn{5}{c}{GeL}\\

 & $Q$ (b)    &RI &E$_{7 \rightarrow 6}$ (keV)  &A  (1/keV)  & $\chi_{red}^2$  & $Q$ (b)  &RI &E$_{7 \rightarrow 6}$ (keV) &A (1/keV)  & $\chi_{red}^2$   \\
\hline
Full \footnotemark[1]                           & 1.97(2) & 0.118(7)     & 359.8(1) &$\leq$10$^{-11}$  & 1.72  & 1.93(5) & 0.17(1)&359.9(1) &0.0043(6)  &1.25  \\  
no $5f_{5/2} \rightarrow 4d_{5/2}$      & 1.98(2) & 0.118(7)     & 359.8(1) &$\leq$10$^{-10}$  & 1.62  & 1.96(4) & 0.168(9) &359.9(1) &0.0044(5)  &1.22  \\  
no $5g_{7/2} \rightarrow 4f_{7/2}$      & 2.04(3) & 0.099(8)     & 359.7(1) &0.006(5)  & 3.07      & 2.05(5) & 0.155(14)&359.8(1) &0.0050(6) &1.92  \\
no weak transitions \footnotemark[2]  & 2.05(2) & 0.097(7)     & 359.7(1) &0.0001(3)  & 2.99    & 2.07(5) & 0.154(9) &359.8(1) &0.0051(6)  &1.86  \\
fix $5g/5f$ population\footnotemark[3]   & 1.90(2) & 0.170          & 359.8(1) &0.0013(5)  & 2.22    & 1.99(8) & 0.118      &359.9(1) &0.0026(5) & 1.62  \\
 
\end{tabular}
\end{ruledtabular}
\footnotetext[1]{Full fit as described in Section~\ref{sec:hfsanalysis}.}
\footnotetext[2]{The $5f_{5/2} \rightarrow 4d_{5/2}$ and $5g_{7/2} \rightarrow 4f_{7/2}$ multiplets are removed from the fit.}
\footnotetext[3]{Relative intensity fixed to 0.17 for GeR and 0.12 for GeL.}

\end{table*}
%%%%%%%%%%%%%%%%%%%%%%%%%%%%%%%%%%%%%%%%%%%%%%

%----------------------------------------------------------------------------------------------------------------------------------------------------
\subsection{\label{sec:discussion}Quadrupole moments and uncertainty}
The values of quadrupole moments with their statistical errors are collected in the
Tables~\ref{tab:quadrupole1} and \ref{tab:quadrupole2}. 
To evaluate possible systematic errors of different parameters, like the line shape, the $\sigma$ of the experimental line shape, 
and the description of the background, their influence on the extracted value of the quadrupole moments was studied separately 
in a systematic way and is reported in Table~\ref{tab:quadrupolesyst} for the two targets. 
The effects were checked for small variations of the
$\chi ^2$ with respect to the values reported in the Tables~\ref{tab:quadrupole1} and \ref{tab:quadrupole2}. \\
The effect of variation in the modeling of the background and the line shape turned out to be negligible
with respect to the value of the quadrupole moment. 
The sensitivity of our results to the assumed background
was examined by comparing the hfs parameters obtained with our constant background model with those using a linear 
or quadratic form of the background.\\ 
The influence of the experimental line shape was investigated by sampling the parameters describing the
line shapes. Out of 1000 sampled line shapes, around 300 could simultaneously fit the shape of the background lines
with reasonable values of $\chi ^2$.
Each of these line shapes was then used in the fit of the hf complex and the distribution of the extracted  
values of the quadrupole moments was fitted with a Gaussian. 
The centroid of the quadrupole moment distribution showed no variation with respect to the quadrupole
moment given by the best line shape and the sigma of the Gaussian distribution was taken as uncertainty.\\
In a similar way, the effect of the $\sigma$ was checked by sampling the values within its statistical 
uncertainty while leaving fixed the other parameters of the line shape. Also in this case the centroid of the distribution of the 
extracted quadrupole moments showed no variation with respect to the  value of the best line shape 
but with a larger uncertainty. \\
Finally, most sensitive was the relative intensity of the $5g_{9/2} \rightarrow 4f_{7/2}$ versus $5f_{7/2} \rightarrow 4d_{5/2}$
transition. As described in Section~\ref{sec:hfsanalysis} the fits of the two detectors do not converge to 
the same ratio. In Table~\ref{tab:quadrupolesyst} the variation of the quadrupole moment obtained when the ratio 
$5g_{9/2} \rightarrow 4f_{7/2}$ versus $5f_{7/2} \rightarrow 4d_{5/2}$ is fixed to the medium value of the two detectors 
is  reported. This variation has been added in the systematic uncertainty. 

%%%%%%%%%%%%%%%%%%%%%%%%%%%%%%%%%%%%%%%%%%%%%%
%\squeezetable
\begin{table}

\caption{\label{tab:quadrupolesyst} The variation of the extracted quadrupole moment $\Delta Q$ in barn  due to various
systematics effects and its uncertainty is analysed on the data of $^{185}$Re/\ $^{187}$Re. 
In the cells where only one value is reported, the effect is the same for the two isotopes.}
\begin{ruledtabular}
\begin{tabular}{ccccc}
Effect & \multicolumn{2}{c}{GeR}&\multicolumn{2}{c}{GeL}\\
                   & $\Delta Q$ (b)    &error (b)   & $\Delta Q$ (b)  &error (b)    \\
\hline
Bkg model             &0.0      &0.01                  &0.0       &0.01/\ 0.03       \\
Line shape            & 0.0     & 0.01                 & 0.0      & 0.01/\ 0.02    \\  
$\sigma$                  & 0.0     & 0.02/\ 0.03        & 0.0     & 0.07/\ 0.06      \\   
RI   & -0.04/\ -0.03   & 0.04/\ 0.03      &0.03/\ 0.03  & 0.03/\ 0.03     \\
\hline
Total   & -0.04/\ -0.03   & 0.05      &0.03/\ 0.03  & 0.08     \\

\end{tabular}
\end{ruledtabular}

\end{table}
%%%%%%%%%%%%%%%%%%%%%%%%%%%%%%%%%%%%%%%%%%%%%%
The final quadrupole moments with their uncertainty are \\
$^{185}Q$ = 2.07 $\pm$ 0.02 (stat) $\pm$ 0.05 (syst)   \\ 
$^{187}Q$ = 1.94 $\pm$ 0.02 (stat) $\pm$ 0.05 (syst)  \\
and \\
$^{185}Q$ = 2.07 $\pm$ 0.05 (stat) $\pm$ 0.08 (syst)   \\ 
$^{187}Q$ = 1.96 $\pm$ 0.05 (stat) $\pm$ 0.08 (syst) \\
for GeR and GeL, respectively. Given the larger uncertainty in GeL, a combined analysis of the
two detectors is clearly not worthwhile.
The ratio of the quadrupole moments was not fixed in our fits and amounts to 2.07(5)/1.94(5) = 1.067(35) 
in very good agreement with the very precise value of 1.056709(17) reported by S.L. Segel~\cite{Seg78}.\\
The extracted $Q$-values are smaller compared to the values of $^{185}Q$ = 2.21(4)~barn and 
$^{187}Q$ = 2.09(4)~barn reported in Ref.~\cite{Kon81}.  
Two weak multiplets namely $5g_{7/2} \rightarrow 4f_{7/2}$ and $5f_{5/2} \rightarrow 4d_{5/2}$ have  
been introduced in the present analysis which were not included in the previous work.
Their effect on the extracted quadrupole moment is reported in Table~\ref{tab:quadrupole1} and Table~\ref{tab:quadrupole2}.
While the inclusion of the very weak $5f_{5/2} \rightarrow 4d_{5/2}$ does not modify the results of the quadrupole moment, 
the $5g_{7/2} \rightarrow 4f_{7/2}$ multiplet has stronger influence and it leads to a lower value of quadrupole moment 
explaining the discrepancy to the values reported in~\cite{Kon81}.
The addition of the  $5g_{7/2} \rightarrow 4f_{7/2}$ multiplet in the fitting of the hfs was necessary in order to 
properly reproduce the rising slope at low energy of the experimental spectrum as can be inferred by the 
significantly higher value of reduced $\chi^2$ obtained when this 
transition is removed from the fit. This effect clearly shows that the isotopically pure muonic X-ray spectra 
could be sensitive to transitions of relative intensity of only a few \%. Since the fitted hfs spectrum is not reported 
in~\cite{Kon81} neither are the values of $\chi^2$, we cannot judge the quality of the fit 
and consequently the sensitivity of that experimental spectrum to weaker transitions.

\section{Conclusions}
The hfs of the $5g \rightarrow 4f$ X-ray transition in muonic $^{185,187}$Re has been investigated.
The extracted values of the quadrupole moments have been determined based on high-quality isotopically 
pure muonic X-ray spectra of $^{185,187}$Re and state-of-the-art theoretical calculations and fitting procedures.
The quadrupole moments $Q$~=~2.07(5)~barn and  $Q$~=~1.94(5)~barn are measured for $^{185,187}$Re, respectively. 
The disagreement with values in literature extracted with the same procedure has been understood 
from the higher sensitivity of the muonic X-ray spectra of isotopically pure targets to weak hyperfine transitions. \\
The measurement of the hyperfine splitting of muonic X rays allows the extraction of the quadrupole moment of the nucleus
to a rather high precision compared to the hyperfine splitting in electronic systems
because they do not suffer from the uncertainty in the calculation of a multi-electron system
for the determination of the electric field gradient at the nucleus and the polarisation 
of the electron core. Nevertheless, we have pointed out that
particular care has to be taken in the estimation of the systematic errors  
for what concerns the description of the detector response and the relative intensity of the muonic transitions.\\
This work is part of the muX project which currently pursues at PSI the possibility to extend muonic atom
spectroscopy to elements available in microgram quantities, with a special emphasis on $^{226}$Ra. \\

\begin{acknowledgments}
We gratefully thank L. Simons for many valuable discussions.
This work was supported by the Paul Scherrer Institut through the Career Return Programme, by the Swiss
National Science Foundation through the Marie Heim-V\"ogtlin grant No. 164515 and the project grant No.
200021\_165569, by the Cluster of Excellence "Precision Physics, Fundamental Interactions, and Structure of Matter"
(PRISMA EXC 1098 and PRISMA+ EXC 2118/1) funded by the German Research Foundation (DFG) 
within the German Excellence Strategy (Project ID 39083149). FW has been supported by the 
German Research Foundation (DFG) under Project WA 4157/1. 
Most of the theory results in this article are part of the PhD thesis work of NM, which was published at the Heidelberg University, Germany.
The experiment was performed at the $\pi$E1 beam line of PSI. We would like to thank the
accelerator and support groups for the excellent conditions. Technical support by F. Barchetti, F. Burri, M. Meier
and A. Stoykov from PSI and B. Zehr from the IPP workshop at ETH Z\"urich is gratefully acknowledged. 
\end{acknowledgments}

%\nocite{*}

\bibliography{mybib}% Produces the bibliography via BibTeX.

\end{document}